\documentclass[intlimits,twoside,a4paper]{article}

\usepackage{amsmath,amssymb}
\usepackage{graphicx}

\usepackage[T2A]{fontenc}
\usepackage[cp1251]{inputenc}

\providecommand{\abs}[1]{\lvert#1\rvert}
\DeclareMathOperator{\Imp}{Im}

\allowdisplaybreaks[3]

\newcommand{\rmi}{\mathrm{i}}%

\usepackage[eqsecnum]{cmpj2}

\issue{2015}{18}{4}{43004}
\doinumber{10.5488/CMP.18.43004}

\title[Phonon-like excitations in the two-state Bose-Hubbard model]
{Phonon-like excitations in the two-state Bose-Hubbard model}
\author[I.V. Stasyuk, O.V. Velychko, O. Vorobyov]{I.V. Stasyuk, O.V. Velychko, O. Vorobyov}
\address{Institute for Condensed Matter Physics
of the National Academy of Sciences of Ukraine, \\
1~Svientsitskii~St., 79011 Lviv, Ukraine}

\date{Received October 5, 2015, in final form November 5, 2015}
\authorcopyright{I.V. Stasyuk, O.V. Velychko, O. Vorobyov, 2015}

\begin{document}

\maketitle

\begin{abstract}
The spectrum of phonon-like collective excitations in the system
of Bose-atoms in optical lattice (more generally, in the system of
quantum particles described by the Bose-Hubbard model) is
investigated. Such excitations appear due to displacements of
particles with respect to their local equilibrium positions. The
two-level model taking into account the transitions of bosons
between the ground state and the first excited state in potential
wells, as well as interaction between them, is used. Calculations
are performed within the random phase approximation in the
hard-core boson limit. It is shown that excitation spectrum in
normal phase consists of the one exciton-like band, while in the
phase with BE condensate an additional band appears. The
positions, spectral weights and widths of bands strongly depend on
chemical potential of bosons and temperature. The conditions of
stability of a system with respect to the lowering of symmetry and
displacement modulation are discussed.
\pacs 03.75.Hh, 03.75.Lm, 64.70.Tg, 71.35.Lk, 37.10.Jk, 67.85.-d
\end{abstract}

\section{Introduction}

Bose-Hubbard model (BHM) \cite{wrk01,Jaksch02} is a well known
model in the solid state physics. In its most applications it is
used to describe the thermodynamics and energy spectrum of
ultracold bosonic atoms in optical lattices. The main attention
was usually paid to phase transition between normal (NO) phase and
phase with the Bose-Einstein (BE) condensate [so-called
Mott-insulator (MI)~--- superfluid state (SF) transition]
\cite{wrk03,wrk04,wrk05,wrk06}. The model is also intensively used
for theoretical description of other phenomena, such as quantum
delocalization of hydrogen atoms adsorbed on the surface of
transition metal \cite{Astaldi07,Nishijima08}, quantum diffusion
of light particles on the surface or in the bulk
\cite{Reilly09,Ignatyuk10}, thermodynamics of the impurity ion
intercalation into semiconductors \cite{Velychko11,Mysakovych12}.

As a general rule, the theoretical consideration is devoted to the
behaviour of atoms confined in the lowest vibrational levels in
the potential wells of a lattice. However, the study of a quantum
delocalization or diffusion reveals an important role of excited
vibrational states of particles (ions) in localized positions with
a much higher probability of ion hopping between them
\cite{Reilly09,Puska13,Brenig14}. In this connection, a
possibility of BE condensation in the excited Bloch bands in
optical lattices is also considered; in this case, the condition
of their sufficient occupation due to the optical pumping (see,
e.g., \cite{Muller15}) is imposed. The Bose-Hubbard model was
extended to higher vibrational bands and, in the framework of such
a generalization, the MI-SF transition to the phase with BE
condensate in the pumping-induced quasi-equilibrium state of the
system has been described \cite{Scarola16}. In the case of orbital
degeneracy of  the excited state (the p-state, for example) and
anisotropy of hopping parameters, the conception of multi-flavour
bosons (that correspond to variously polarized bands) was
introduced \cite{Isacsson17}. The possibility of appearance of
unconventional p-orbital BE condensate with the non-zero
incommensurate wave vector was shown \cite{Cai18,Liu19}. The
double-well lattices open up a new field of researches in this
direction \cite{Stojanovii20}.

Contrary to that, in the equilibrium case, the BE condensation
involving the excited states was not sufficiently studied in the
framework of ordinary Bose-Hubbard model. One can mention the
systems of spin-1 bosons where the multiplets of local states form
the closely-spaced excited levels. It was shown in
\cite{Kimura21,Pai22} that MI-SF transition can be of the first
order when a single-site spin interaction is of the
antiferromagnetic type. A similar effect also takes place for
multicomponent Bose system in the optical lattice \cite{Chen23}.

A change of the phase transition (PT) order (from the 2nd to the
1st order) is also possible in the equilibrium lattice gas of
bosons with the transfer of particles over the excited states. We
have considered such a problem in \cite{Stasyuk24,wrk25,wrk26}
using the Bose-Hubbard model where the only excited nondegenerated
state on the lattice site besides the ground state (the so-called
two state BHM) was taken into account. The model corresponds to 1D
or strongly anisotropic (quasi-1D) optical lattice; it is also
close to the situation in a system of light particles (protons,
lithium ions) adsorbed on the metal surface.

In the present work we continue the investigation started in
\cite{Stasyuk24}. In addition to the analysis of the boson
one-particle spectrum performed in \cite{Stasyuk24,wrk25} we study
the collective dynamics of bosons. The pair interaction between
particle displacements with respect to their equilibrium positions
in the neighbouring wells will be taken into account. In a
harmonic case, such displacements are expressed in quantum
language in terms of transitions between the nearest vibrational
levels. Due to the above mentioned interaction, the collective
modes appear; their energy is a function of the wave vector.
Attention is paid to changes in the spectrum after the MI-SF
transition (both of the 2nd and 1st order) and to the BE
condensate appearance. We perform our consideration in the
hard-core boson (HCB) limit, which means no more than one particle
per site (the single-site problem is here a three-level one). Such
a model of hard-core bosons is well known; it is used in a  wide
range of problems, e.g., local electron pairing mechanism of
superconductivity \cite{Micnas27a}, ion transport in ionic
conductors \cite{Mahan28,Stasyuk29}, BE condensation and related
phenomena in optical lattices \cite{Sengupta30,wrk31,Stasyuk32}.

\section{Model}

To consider the phonon-type dynamics in the Bose-atom system in an
optical lattice, we use a model which is a simple generalization
of the hard-core boson model. Describing vibrational states of
Bose-atom in a separate potential well in the lattice, we take
into account the first excited level besides the ground one. Let
us suppose that this level is nondegenerate; such a situation can
be realized in the case of low local symmetry. At the same time,
we impose a usual (for hard-core bosons) restriction: no more than
one particle per lattice site. The model of this type was used by
us while investigating the conditions, at which the Bose-Einstein
(BE) condensate appears when the particles are hopping only over
the excited states. Hamiltonian of the model was written as
follows:
\begin{align}
    \hat{H} &= \sum_{ij} t(i,j)b_i^+ b_j
        + {\sum}_{ij} t'(i,j)c_i^+ c_j
        + \sum_{ij} t''(i,j)(b_{i}^{+}c_{j}+c_{i}^{+}b_{j})
        \notag\\
    &\quad        + \sum_i(\varepsilon-\mu) b_i^+b_{i}
    + \sum_{i}(\varepsilon'-\mu) c_i^+c_{i} ,
    \label{eq2-01}
\end{align}
where $b_{i}$ and $b_{i}^{+}$ ($c_{i}$ and $c_{i}^{+}$) are Bose
operators of annihilation and creation of particles in the ground
(excited) states; $\varepsilon$ and $\varepsilon'$ are respective
energies of states ($\delta=\varepsilon'-\varepsilon$ is the
on-site excitation energy); $\mu$ is the boson chemical potential.
For matrix elements describing the particle hoppings onto the
neighbouring sites we take \cite{Stasyuk24}
\begin{equation}
    t(i,j)=0,\qquad
    t''(i,j)=0,\qquad
    t'(i,j)\neq0
    \label{eq2-02}
\end{equation}
basing on the  above mentioned arguments. It should be also noted
that such an approximation agrees with the results of numerical
estimates for optical lattice performed by T.~M\"{u}ller (see
\cite{Mueller06}), where it was shown that the tunneling matrix
elements $t$ and $t'$ can significantly differ by up to one order
(or even more) of magnitude.

Introducing the Hubbard operators $X_{i}^{n,m;n'm'}\equiv|i,n,m\rangle\langle
i,n',m'|$, acting on a single-site basis \linebreak $|n_{i}^{b},
n_{i}^{c}\rangle\equiv|i,n_{i}^{b}, n_{i}^{c}\rangle$ (which is formed by
particle occupation numbers in the ground and in the excited states), we can
write
\begin{equation}
    b_{i}
    =
    \sum_{n_{b}=0}^{\infty}\sum_{n_{c}=0}^{\infty}\sqrt{n_{b}+1}X_{i}^{n_{b}, n_{c}; n_{b+1}, n_{c}},
    \qquad
    c_{i}
    =
    \sum_{n_{b}=0}^{\infty}\sum_{n_{c}=0}^{\infty}\sqrt{n_{c}+1}X_{i}^{n_{b}, n_{c}; n_{b}, n_{c+1}}.
    \label{eq2-03}
\end{equation}
After transition to the hard-core boson limit, there remain only
the $|0,0\rangle$, $|1,0\rangle$ and $|0,1\rangle$ states. In this
case the relations (\ref{eq2-03}) take the following form:
\begin{equation}
b_{i}=X_{i}^{00;10},\qquad c_{i}=X_{i}^{00;01}.
\label{eq2-04}
\end{equation}
Respectively,
\begin{equation}
b_{i}^{+}=X_{i}^{10;00},\qquad c_{i}^{+}=X_{i}^{01;00}.
\label{eq2-05}
\end{equation}
In what follows, we use the shortened notations
\begin{equation}
    |0,0\rangle=|0\rangle,\qquad
    |1,0\rangle=|1\rangle,\qquad
    |0,1\rangle=|2\rangle.
    \label{eq2-06}
\end{equation}
In such a case,
\begin{equation}
    b_{i}=X_{i}^{01},\qquad b_{i}^{+}=X_{i}^{10};
    \qquad
    c_{i}=X_{i}^{02},\qquad c_{i}^{+}=X_{i}^{20}.
    \label{eq2-07}
\end{equation}

When particles are moving through the lattice only in the excited
state, BE condensate is described by the order parameter $\langle
c_{i}\rangle=\langle c_{i}^{+}\rangle=\xi$ (or $\langle
X_{i}^{02}\rangle=\langle X_{i}^{20}\rangle=\xi$). With the help
of identity
\[
c_{i}^{+}c_{j}=(c_{i}^{+}+c_{j})\xi+(c_{i}^{+}-\xi)(c_{j}-\xi)-\xi^{2}.
\]
Hamiltonian (\ref{eq2-01}) can be presented as follows:
\begin{equation}
\hat{H}=\hat{H}_{0}+\sum_{ij}t'(i,j)(c_{i}^{+}-\xi)(c_{j}-\xi)+N\abs{t'_0}\xi^{2},
\label{eq2-08}
\end{equation}
where the first term $\hat{H}_{0}$ is the mean-field Hamiltonian
\begin{equation}
    \hat{H}_{0}
    =
    -\abs{t'_0}\xi\sum_{i}(c_{i}^{+}+c_{i})
    +\sum_{i}(\varepsilon-\mu)b_{i}^{+}b_{i}
    +\sum_{i}(\varepsilon'-\mu)c_{i}^{+}c_{i}+N\abs{t'_0}\xi^{2}.
    \label{eq2-09}
\end{equation}
Here, the notation $\sum_{j}t'(i,j)=t'_{0}=-\abs{t'_0}$ is used;
$t'_{0}\equiv t'_{\vec{q}=0}$ is the zero-momentum
Fourier-transform of the hopping matrix element. We consider the
case, when $t'(i,j)<0$ \cite{Stasyuk24}.

In terms of $X$-operators acting on the three-state basis (\ref{eq2-06})
\begin{equation}
    \hat{H}_{0}
    =
    \sum_{i}\hat{H}_{i},
    \qquad
    \hat{H}_{i}
    =
    (\varepsilon-\mu)X_{i}^{11}+(\varepsilon{'}-\mu)X_{i}^{22}-
    \abs{t'_0}\xi(X_{i}^{02}+X_{i}^{20})+\abs{t'_0}\xi^{2}.
    \label{eq2-10}
\end{equation}
The single-site Hamiltonian $\hat{H}_{i}$ is diagonalized by
transformation
\begin{align}
|0\rangle&=\cos\alpha\cdot|\tilde{0}\rangle-\sin\alpha\cdot|\tilde{2}\rangle,\nonumber\\
|1\rangle&=|\tilde{1}\rangle,\nonumber\\
|2\rangle&=\cos\alpha\cdot|\tilde{2}\rangle+\sin\alpha\cdot|\tilde{0}\rangle,
\label{eq2-11}
\end{align}
where
\begin{equation}
    \cos2\alpha=\frac{\varepsilon'-\mu}{\sqrt{(\varepsilon'-\mu)^{2}+4\xi^{2}t_{0}^{\prime2}}},\qquad
    \sin2\alpha=\frac{2\xi\abs{t'_0}}{\sqrt{(\varepsilon'-\mu)^{2}}+4\xi^{2}t_{0}^{\prime2}}.
    \label{eq2-12}
\end{equation}
As a result,
\begin{equation}
    \hat{H}_{i}
    =
    \tilde{\lambda}_{0}\tilde{X}_{i}^{00}
    +
    \tilde{\lambda}_{1}\tilde{X}_{i}^{11}
    +
    \tilde{\lambda}_{2}\tilde{X}_{i}^{22}+\abs{t'_{0}}\xi^{2},
    \label{eq2-13}
\end{equation}
here, the $\tilde{X}_{i}^{mm}$ operators act on ``tilded'' basis,
and the energy eigenvalues are as follows:
\begin{equation}
    \tilde{\lambda}_{0,2}
    =
    \frac{\varepsilon'-\mu}{2}
    \mp
    \frac12\sqrt{(\varepsilon'-\mu)^{2}+4\xi^{2}t_{0}^{'2}},
    \qquad
    \tilde{\lambda}_{1}
    =
    \varepsilon-\mu.
    \label{eq2-14}
\end{equation}

We intend to investigate the spectrum of phonon-type excitations,
which are connected with displacements of Bose-particles from
their equilibrium positions in the sites of optical lattice. In
the  quantum description, the operators $\tilde{X}_{i}$ of  such
displacements are characterized by their matrix elements. The
latter are nonzero when calculated between the states having
different parity (it is assumed usually that  local potential
wells are harmonic and, correspondingly, the oscillatory wave
functions form a basis of local states). In the case when only one
excited state is taken into account, only matrix elements
$\langle0|\hat{x}|1\rangle$ and $\langle1|\hat{x}|0\rangle$ are of
current interest. In the second quantization representation
\begin{equation}
    \hat{x}_{i}
    =
    d(c_{i}^{+}b_{i}+b_{i}^{+}c_{i})=d\left(X_{i}^{21}+X_{i}^{12}\right),
    \label{eq2-15}
\end{equation}
here, $d=\langle0|\hat{x}|1\rangle=\langle1|\hat{x}|0\rangle$

Collective vibrations arise due to interaction between particle
displacements. Let us write it in the following form:
\begin{equation}
\hat{H}'=\frac12\sum_{ij}\Phi(i,j)\hat{x}_{i}\hat{x}_{j}.
\label{eq2-16}
\end{equation}
As distinct from direct interparticle interaction (that looks like
$~V(i,j)n_{i}n_{j}$), the interaction (\ref{eq2-16}) has another
nature; it is caused by transitions between the ground and excited
states. This interaction can be considered as an analogue of the
so-called resonant interaction that is responsible for the
dynamics of excitations (Frenkel excitons) in molecular crystals.

In terms of ``tilded'' operators
\begin{align}
    b_{i}&=\cos\alpha\cdot\tilde{X}_{i}^{01}-\sin\alpha\cdot\tilde{X}_{i}^{21},\nonumber\\
    c_{i}&=\sin\alpha\cos\alpha\cdot\left(\tilde{X}_{i}^{00}-\tilde{X}_{i}^{22}\right)+\cos^{2}\alpha\cdot\tilde{X}_{i}^{02}
    -\sin^{2}\alpha\cdot\tilde{X}_{i}^{20},
    \label{eq2-17}
\end{align}
and
\begin{equation}
\hat{x}_{i}=d\sin\alpha\cdot\left(\tilde{X}_{i}^{10}+\tilde{X}_{i}^{01}\right)+d\cos\alpha\cdot
\left(\tilde{X}_{i}^{12}+\tilde{X}_{i}^{21}\right).
\label{eq2-18}
\end{equation}

We now use the technique of the Zubarev two-time temperature
Green's functions to study the vibrational spectrum. Let us
introduce the commutator Green's function of displacements
\begin{equation}
\langle\langle\hat{x}_{i}|\hat{x}_{j}\rangle\rangle
=d^{2}\langle\langle(c_{i}^{+}b_{i}+b_{i}^{+}c_{i}|(c_{j}^{+}b_{j}+b_{j}^{+}c_{j})\rangle\rangle.
\label{eq2-19}
\end{equation}
Using (\ref{eq2-18}), this function can be presented as  linear
combination of functions defined on $\tilde{X}$-operators
\begin{align}
    \langle\langle \hat{x}_{i}|\hat{x}_{j}\rangle\rangle
    &=
    d^{2}
    \left[
        \sin^{2}\alpha
        \left(
            G_{ij}^{10,10}+G_{ij}^{01,10}+G_{ij}^{10,01}+G_{ij}^{01,01}
        \right)
    \right.
    \notag\\
    &\quad+
    \cos^{2}\alpha
        \left(
            G_{ij}^{12,12}+G_{ij}^{21,12}+G_{ij}^{12,21}+G_{ij}^{21,21}
        \right)
    \notag\\
    &\quad+
        \sin\alpha\cos\alpha
        \left(
            G_{ij}^{12,10}+G_{ij}^{21,10}+G_{ij}^{12,01}+G_{ij}^{21,01}
        \right.
    \notag\\
    &
    \left.\!
        \left.\!
        \quad{}+
            G_{ij}^{10,12}+G_{ij}^{01,12}
            +
            G_{ij}^{10,21}+G_{ij}^{01,21}
        \right)
    \right],
    \label{eq2-20}
\end{align}
where
\begin{equation}
    G_{ij}^{pq,rs}
    =
    \langle\langle
        \tilde{X}_{i}^{pq}|\tilde{X}_{j}^{rs}
    \rangle\rangle.
    \label{eq2-21}
\end{equation}
The poles of function (\ref{eq2-18}) determine the energies of collective
vibrations while the imaginary part
$\Imp
\langle\langle\hat{x}_{i}|\hat{x}_{j}\rangle\rangle_{\omega+\rmi\varepsilon}$
is connected with spectral density
\begin{equation}
\rho(\omega)=-\frac{2}{\hbar}\Imp\langle\langle\tilde{x}_{i}|\tilde{x}_{i}\rangle\rangle_{\omega+\rmi\varepsilon}
\label{eq2-22}
\end{equation}
which provides a more complete description of the vibrational
spectrum, providing in particular information on the statistical
weights of boson vibrational modes.

\section{Green's function of displacements}

We calculate the Green's function (\ref{eq2-21}) in the random
phase approximation (RPA) using the Hamiltonian
\begin{equation}
    \tilde{H}=\sum_{i}\hat{H}_{i}+\hat{H}',
    \label{eq3-01}
\end{equation}
where we add the pair interaction between the particle
displacements to the mean-field Hamiltonian. Let us employ a
standard scheme of the equation of motion method. Accordingly, in
the frequency representation, the following equation
\begin{equation}
\hbar \omega\langle\langle\tilde{X}_{i}^{pq}|\tilde{X}_{j}^{rs}\rangle\rangle
=\frac{\hbar}{2\pi}\delta_{ij}\langle\tilde{X}^{ps}\delta_{qr}-\tilde{X}^{rq}\delta_{sp}\rangle
+\langle\langle[\tilde{X}_{i}^{pq},\tilde{H}]|\tilde{X}_{j}^{rs}\rangle\rangle
\label{eq3-02}
\end{equation}
for Green's function (\ref{eq2-21}) is used.

Equation of motion for separate operator $\tilde{X}_{i}^{pq}$ can
be  easily written taking into account the commutation relations
for $\tilde{X}$-operators. In particular, for $\tilde{X}^{1r}_{l}$
operator, we have
\begin{align}
    [\tilde{X}_{l}^{12},\hat{H}]
    &=
    \left[(\varepsilon'-\mu)\cos^{2}\alpha-(\varepsilon-\mu)\right]\tilde{X}_{l}^{12}
    +(\varepsilon'-\mu)\sin\alpha\cos\alpha\tilde{X}_{l}^{10}\nonumber\\
    &\quad+
    \sum_{j}t'(l,j)\left(\cos^{2}\alpha\tilde{X}_{l}^{10}-\sin\alpha\cos\alpha\tilde{X}_{l}^{12}\right)c_{j}\nonumber\\
    &\quad-
    \sum_{i}t'(i,l)c_{i}^{+}\left(\sin^{2}\alpha\tilde{X}_{l}^{10}+\sin\alpha\cos\alpha\tilde{X}_{l}^{12}\right)\nonumber\\
    &\quad+
    \sum_{j}\Phi(l,j)\cos\alpha\left(\tilde{X}_{l}^{11}-\tilde{X}_{l}^{22}\right)\hat{x}_{j}d.
    \label{eq3-03}
\end{align}
In the spirit of RPA scheme, we decouple the pair products of
operators taking into account that mean values of nondiagonal
$\tilde{X}$-operators are equal to  zero in the applied
approximation
\begin{equation}
\tilde{X}_{l}^{10}c_{j}\approx\langle\tilde{X}_{l}^{10}\rangle c_{j}+\tilde{X}_{l}^{10}\langle c_{j}\rangle-\langle\tilde{X}_{l}^{10}\rangle
\langle c_{j}\rangle=\tilde{X}_{l}^{10}\langle c_{j}\rangle,
\label{eq3-04}
\end{equation}
and, similarly,
\begin{equation*}
\tilde{X}_{l}^{12}c_{j}\approx\tilde{X}_{l}^{12} \langle c_{j}\rangle,\qquad
c_{i}^{+}\tilde{X}_{l}^{10}\approx\langle c_{i}^{+}\rangle\tilde{X}_{l}^{10},\qquad
 c_{i}^{+}\tilde{X}_{l}^{12}\approx\langle{c}_{i}^{+}\rangle \tilde{X}_{l}^{10}.
\end{equation*}
At the same time,
\begin{align}
\left(\tilde{X}_{l}^{11}-\tilde{X}_{l}^{22}\right)\tilde{x}_{j}&=\left(\tilde{X}_{l}^{11}-\tilde{X}_{l}^{22}\right)d
\left[\sin\alpha\left(\tilde{X}_{j}^{10}+\tilde{X}_{j}^{01}\right)
+\cos\alpha\left(\tilde{X}_{j}^{12}+\tilde{X}_{j}^{21}\right)\right]\nonumber\\
&\approx\langle\tilde{X}_{l}^{11}-\tilde{X}_{l}^{22}\rangle\hat{x}_{j}.
\label{eq3-05}
\end{align}
Here,
\begin{equation}
    \langle c_{j}\rangle=\sin\alpha\cos\alpha\cdot Q,\qquad
    \langle c_{i}^{+}\rangle=\sin\alpha\cos\alpha\cdot Q,\qquad
    \langle \tilde{X}_{l}^{11}-\tilde{X}_{l}^{22}\rangle=-\sigma,
\label{eq3-06}
\end{equation}
where
\begin{equation}
Q=\langle\tilde{X}^{00}-\tilde{X}^{22}\rangle,\qquad \sigma=\langle\tilde{X}^{22}-\tilde{X}^{11}\rangle
\label{eq3-07}
\end{equation}
(here we consider the uniform case).

As a result,
\begin{equation}
[\tilde{X}_{l}^{12},\hat{H}]=\Delta_{21}\tilde{X}_{l}^{12}+\Delta'\tilde{X}_{l}^{10}-\sigma
d^{2}\cos\alpha\sum_{j}\Phi(l,j)\left[\sin\alpha
\left(\tilde{X}_{j}^{10}+\tilde{X}_{j}^{01}\right)+\cos\alpha\left(\tilde{X}_{j}^{12}+\tilde{X}_{j}^{21}\right)\right].
\label{eq3-08}
\end{equation}
Here,
\begin{align}
\Delta_{21}&=(\varepsilon'-\mu)\cos^{2}\alpha-(\varepsilon-\mu)+\frac12\sin^{2}2\alpha
\cdot Q\cdot\abs{t'_{0}}\equiv\tilde{\lambda}_{2}-\tilde{\lambda}_{1},\nonumber\\
\Delta'&=(\varepsilon'-\mu)\sin\alpha\cos\alpha-\frac12\sin2\alpha\cos2\alpha\cdot Q\cdot\abs{t'_{0}}=0.
\label{eq3-09}
\end{align}
In a similar way, we can come to the equation of motion for the
$\tilde{X}_{l}^{10}$ operator
\begin{equation}
[\tilde{X}_{l}^{10},\hat{H}]=\Delta_{01}\tilde{X}_{l}^{10}+\Delta'\tilde{X}_{l}^{12}-Q'd\sin\alpha\sum_{j}\Phi(l,j)
\left[\sin\alpha\left(\tilde{X}_{j}^{10}+\tilde{X}_{j}^{01}\right)+\cos\alpha\left(\tilde{X}_{j}^{12}+\tilde{X}_{j}^{21}\right)\right],
\label{eq3-10}
\end{equation}
where
\begin{equation}
\Delta_{01}=(\varepsilon'-\mu)\sin^{2}\alpha-(\varepsilon-\mu)-\frac12\sin^{2}2\alpha\cdot Q\cdot\abs{t'_{0}}\equiv\tilde{\lambda}_{0}-\tilde{\lambda}_{1}.
\label{eq3-11}
\end{equation}

Equations of motion for $\tilde{X}_{l}^{21}$ and $\tilde{X}_{l}^{10}$
operator can be  found from (\ref{eq3-08}) and (\ref{eq3-10}) with the help
of relations
\begin{equation}
[\tilde{X}_{l}^{21},\hat{H}]=-[\tilde{X}_{l}^{12},\hat{H}]^{+},\qquad
[\tilde{X}_{l}^{01},\hat{H}]=[\tilde{X}_{l}^{10},\hat{H}]^{+}.
\label{eq3-12}
\end{equation}

Using the formulae (\ref{eq3-08}), (\ref{eq3-10}) and
(\ref{eq3-12}), one can write a set of equations (\ref{eq3-02}) in
the explicit form. In a matrix representation, after the Fourier
transition to wave vectors, it has the following form:
\begin{align}
\|\hat{W}\|\times\left(\begin{array}{c}
G^{12,rs}\\
G^{10,rs}\\
G^{21,rs}\\
G^{01, rs}
\end{array}\right)=\|\hat{I}\|^{rs},
\label{eq3-13}
\end{align}
where
\begin{align}
\|\hat{W}\|=\left(\begin{array}{cccc}
\hbar \omega-\Delta_{21}+\sigma A_{q}& \sigma C_{q}&\sigma A_{q}& \sigma C_{q}\\
Q' C_{q}& \hbar \omega-\Delta_{01}+Q' B_{q}& Q'C_{q} & Q'B_{q}\\
-\sigma A_{q}& -\sigma C_{q}& \hbar \omega+\Delta_{21}-\sigma A_{q}& -\sigma C_{q}\\
-Q' C_{q}& -Q' B_{q}& -Q' C_{q}& \hbar \omega +\Delta_{01}-Q' B_{q}
\end{array}\right),
\label{eq3-14}
\end{align}
and
\begin{equation}
\|\hat{I}\|^{21}=-\frac{\hbar}{2\pi}\left(\begin{array}{c}
\sigma\\
0\\
0\\
0
\end{array}\right),\quad
\|\hat{I}\|^{01}=\frac{\hbar}{2\pi}\left(\begin{array}{c}
0\\
Q'\\
0\\
0
\end{array}\right),\quad
\|\hat{I}\|^{12}=\frac{\hbar}{2\pi}\left(\begin{array}{c}
0\\
0\\
\sigma\\
0
\end{array}\right),\quad
\|\hat{I}\|^{10}=\frac{\hbar}{2\pi}\left(\begin{array}{c}
0\\
0\\
0\\
Q'
\end{array}\right).
\label{eq3-15}
\end{equation}
Here, the notations are introduced
\begin{equation}
A_{q}=d^{2}\Phi_{q}\cos^{2}\alpha,\qquad B_{q}=d^{2}\Phi_{q}\sin^{2}\alpha,\qquad
C_{q}=d^{2}\Phi_{q}\sin\alpha\cos\alpha;
\label{eq3-16}
\end{equation}
moreover,
\begin{equation}
Q'=\langle\tilde{X}^{00}-\tilde{X}^{11}\rangle\equiv Q+\sigma.
\label{eq3-17}
\end{equation}
The function $\Phi_{q}$ is the Fourier-transform of the pair
interaction of displacements.
\[
    \Phi_{q}
    =
    \sum_{ij}{\re}^{\ri\vec{q}(\vec{R}_{i}-\vec{R}_{j})}\Phi(i,j).
\]

A set of equations (\ref{eq3-13}) decomposes into independent
subsets; each of them consists of four equations. Their solutions
can be easily found. After substitution into formula
(\ref{eq2-20}) and after some simplifications, we get
\begin{align}
\langle\langle\hat{x}|\hat{x}\rangle\rangle_{\omega,\vec{q}}&=
-\frac{\hbar}{2\pi}\frac{d^{2}}{D(\omega)}
\left[2\Delta_{01}\sin^{2}\alpha\cdot Q'(\hbar^{2}\omega^{2}-
\Delta_{21}^{2})\right.\nonumber\\
&\left.\quad+2\Delta_{21}\cos^{2}\alpha\cdot \sigma (\hbar^{2}\omega^{2}-\Delta_{01}^{2})\right].
\label{eq3-18}
\end{align}
Denominator $D(\omega)$ is given by an expression
\begin{align}
D(\omega)&=(\hbar^{2}\omega^{2}-\Delta_{21}^{2}) (\hbar^{2}\omega^{2}-\Delta_{01}^{2})+2\sigma A_{q}\Delta_{21}(\hbar^{2}\omega^{2}-\Delta_{01}^{2})
\nonumber\\
&\quad+2Q' B_{q}\Delta_{01}(\hbar^{2}\omega^{2}-\Delta_{21}^{2}).
\label{eq3-19}
\end{align}
Its zeros determine the poles of the Green's function.
Consequently, having solved the equation
\begin{equation}
D(\omega)=0,
\label{eq3-20}
\end{equation}
we can calculate the vibrational spectrum of Bose-particles.

The solutions of equation (\ref{eq3-20}) are as follows:
\begin{equation}
\hbar \omega_{1,2}\equiv\varepsilon_{1,2}=\pm\sqrt{y_{1}},\qquad
\hbar \omega _{3,4}\equiv\varepsilon_{3,4}=\pm\sqrt{y_{2}},
\label{eq3-21}
\end{equation}
where
\begin{align}
y_{1,2}&=\frac12\left(\Delta_{21}^{2}+\Delta_{01}^{2}-2\Delta_{01}Q'B_{q}-2\Delta_{21}\sigma A_{q}\right)\nonumber\\
&\quad\pm\left[\frac14\left(\Delta_{21}^{2}-\Delta_{01}^{2}+2\Delta_{01}Q'B_{q}-2\Delta_{21}\sigma A_{q}\right)^{2}+4\Delta_{01}\Delta_{21}Q'B_{q}
\sigma A_{q}\right].
\label{eq3-22}
\end{align}
In general, there are four branches in spectrum. They form two
pairs which differ by sign among themselves. Their dispersion laws
are determined by the dependence of $A_{q}$, $B_{q}$, and $C_{q}$
functions (which are linear in the $\Phi_{q}$ interaction) upon
the wave vector.

\begin{figure}[!t]
\noindent%
\includegraphics[width=0.45\textwidth]{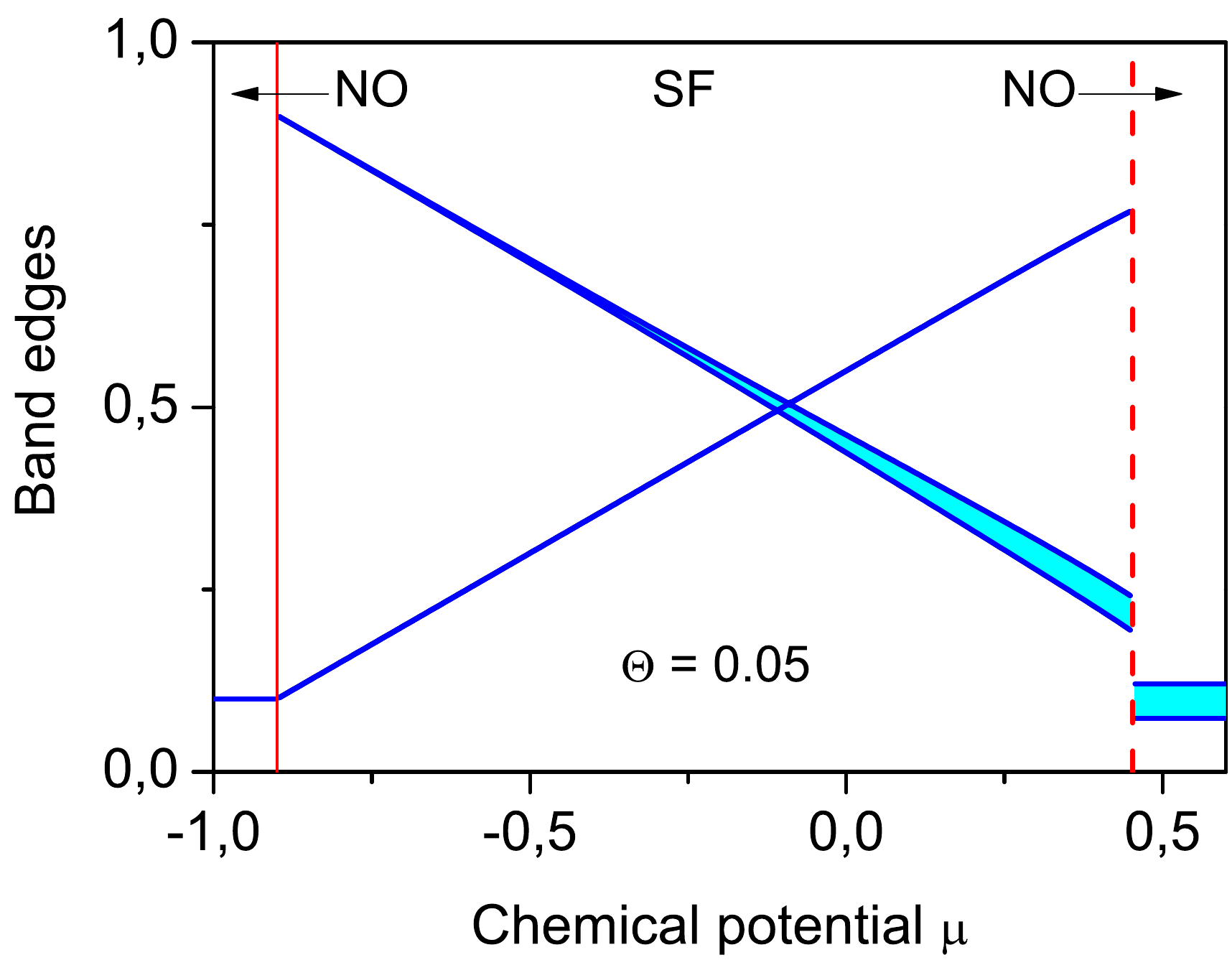}%
\hfill%
\includegraphics[width=0.45\textwidth]{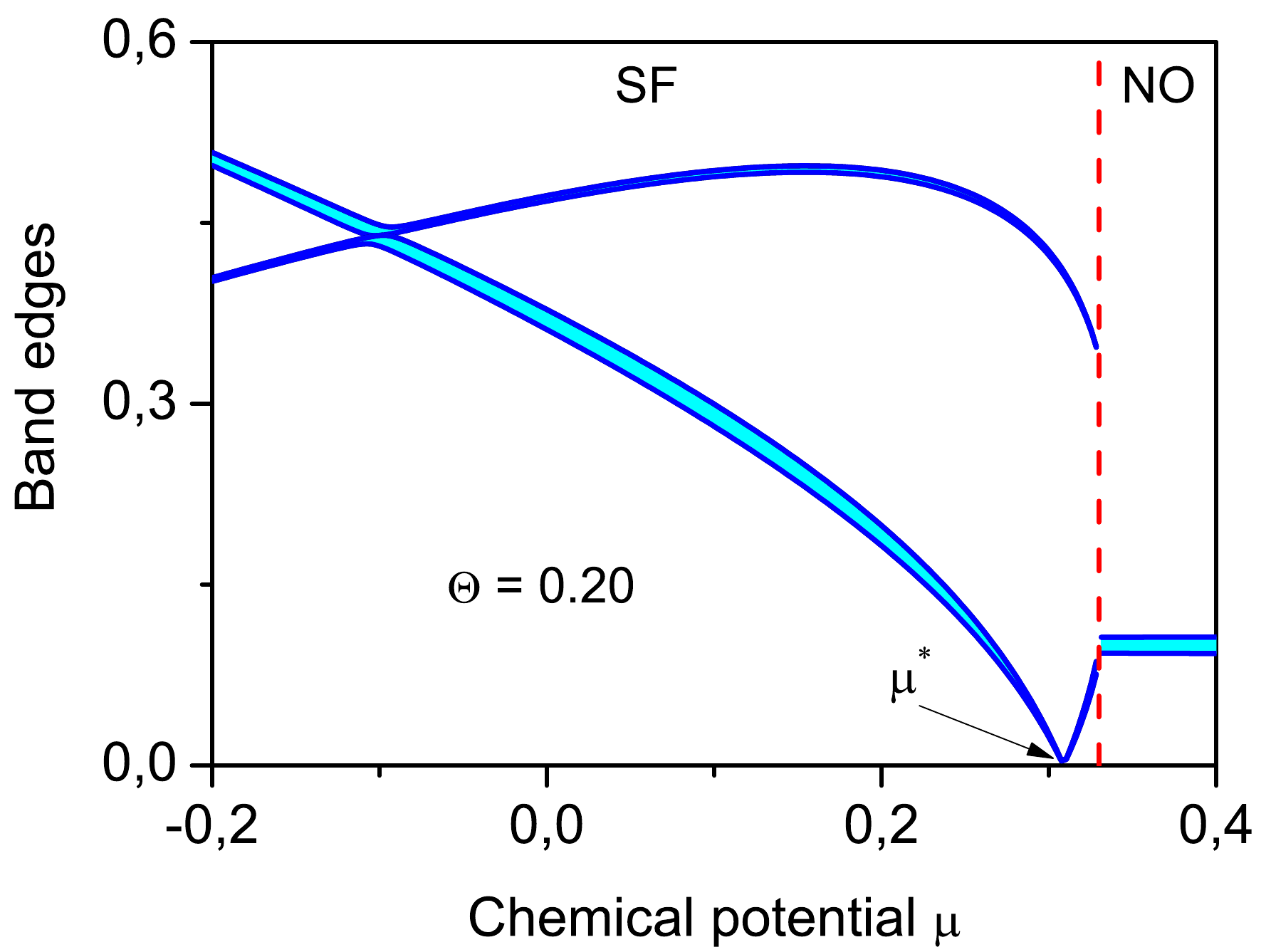}%
\\%
\smallskip
\includegraphics[width=0.448\textwidth]{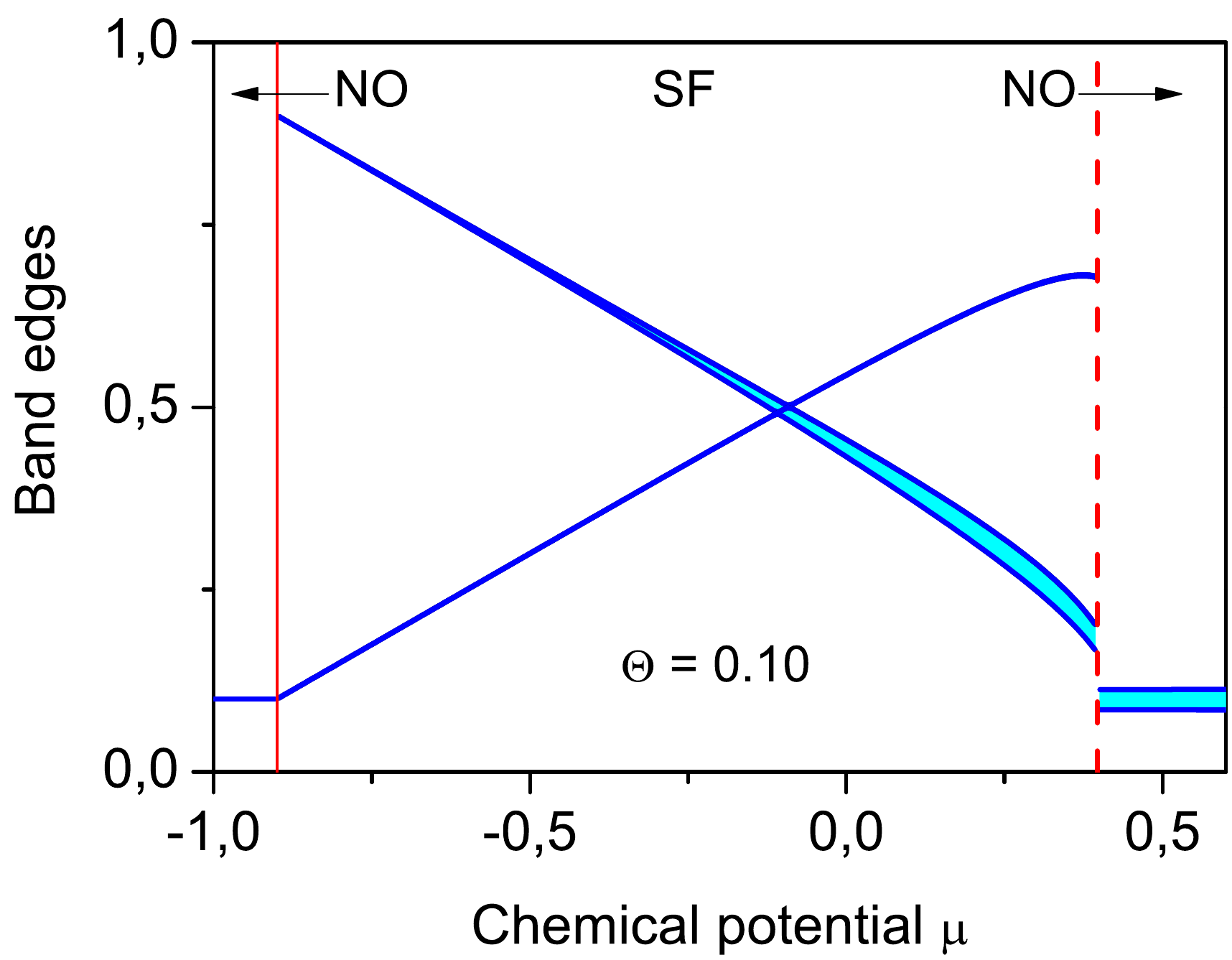}%
\hfill%
\includegraphics[width=0.45\textwidth]{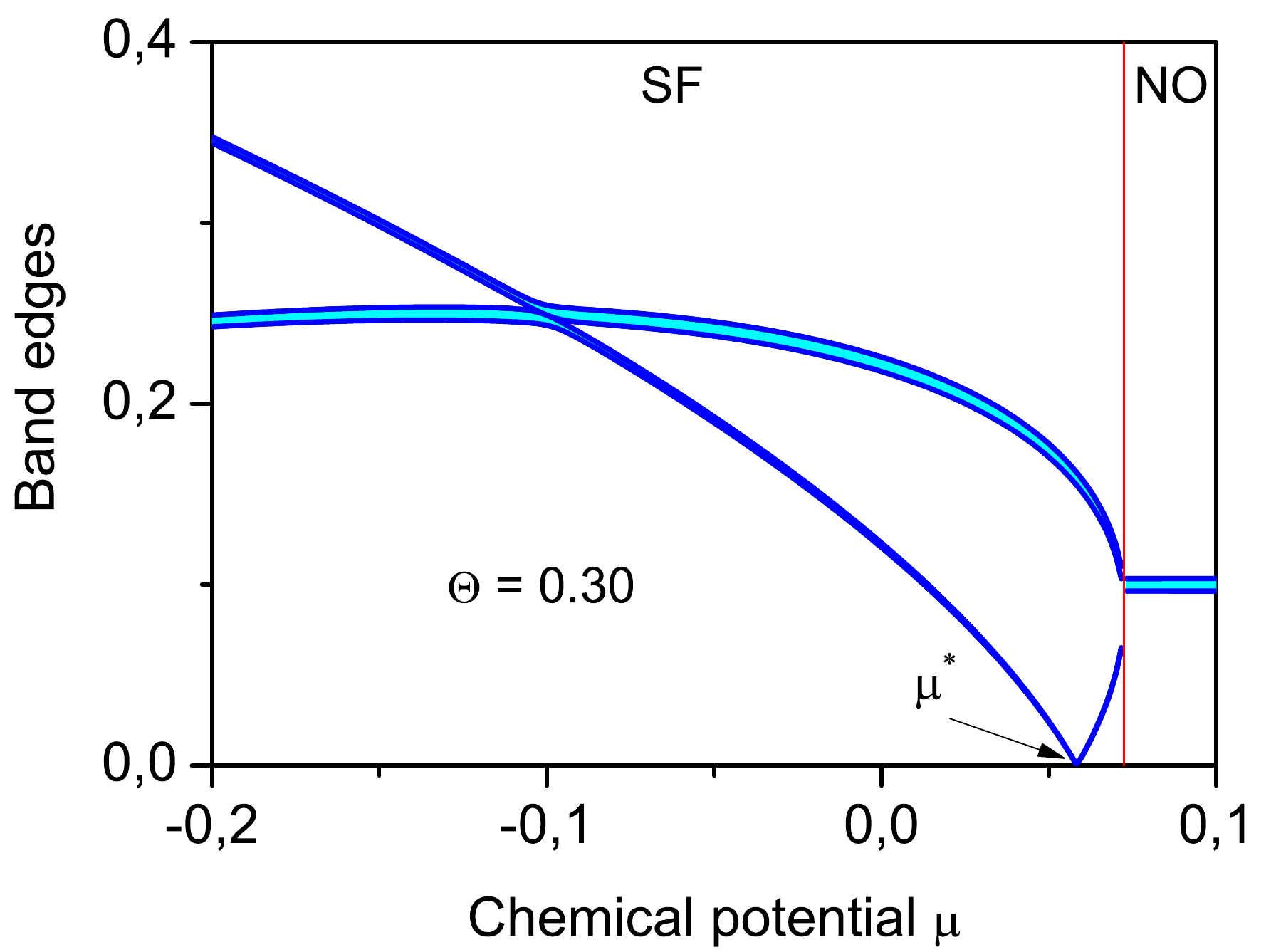}%
\\%
\smallskip
\includegraphics[width=0.45\textwidth]{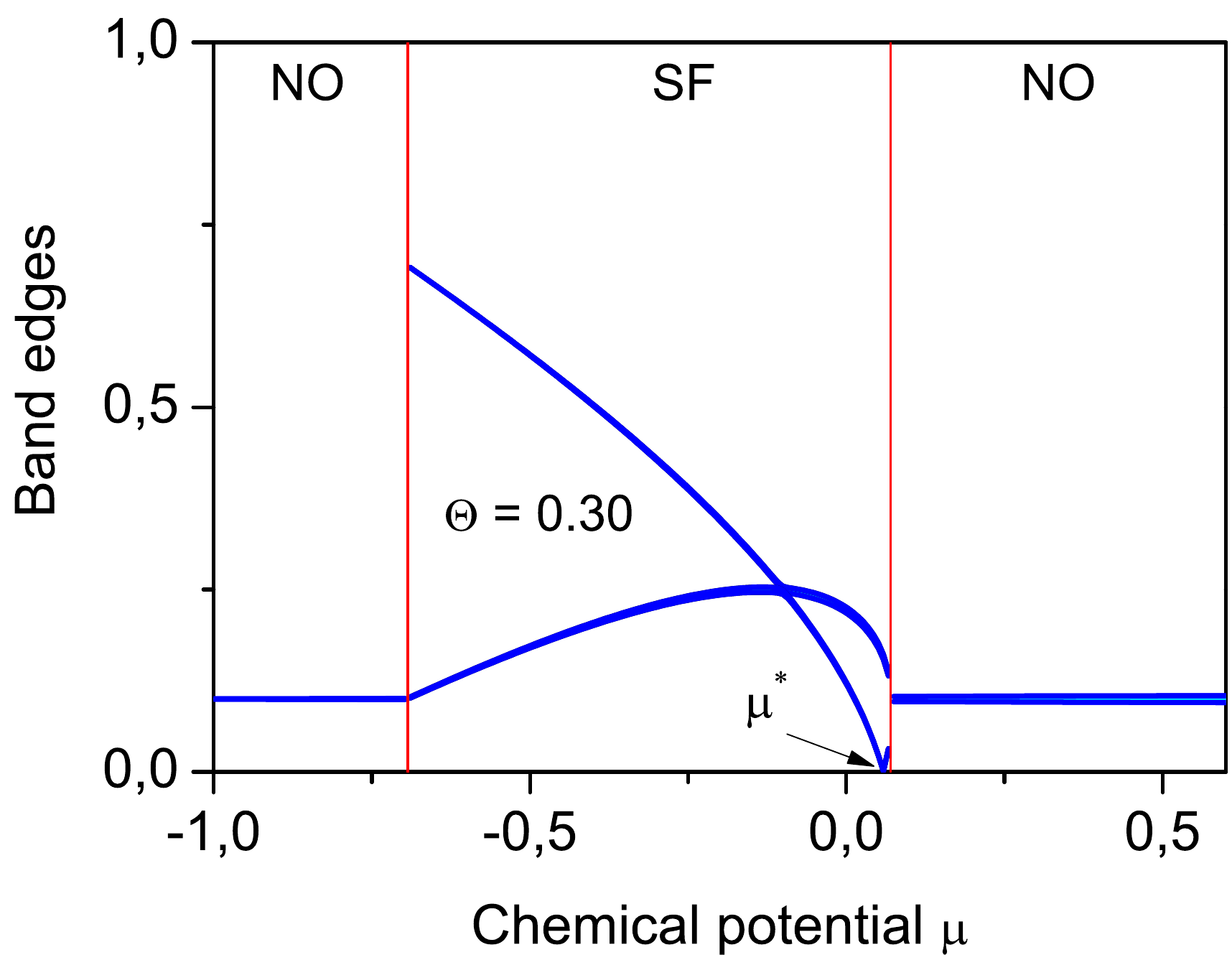}%
\hfill%
\includegraphics[width=0.45\textwidth]{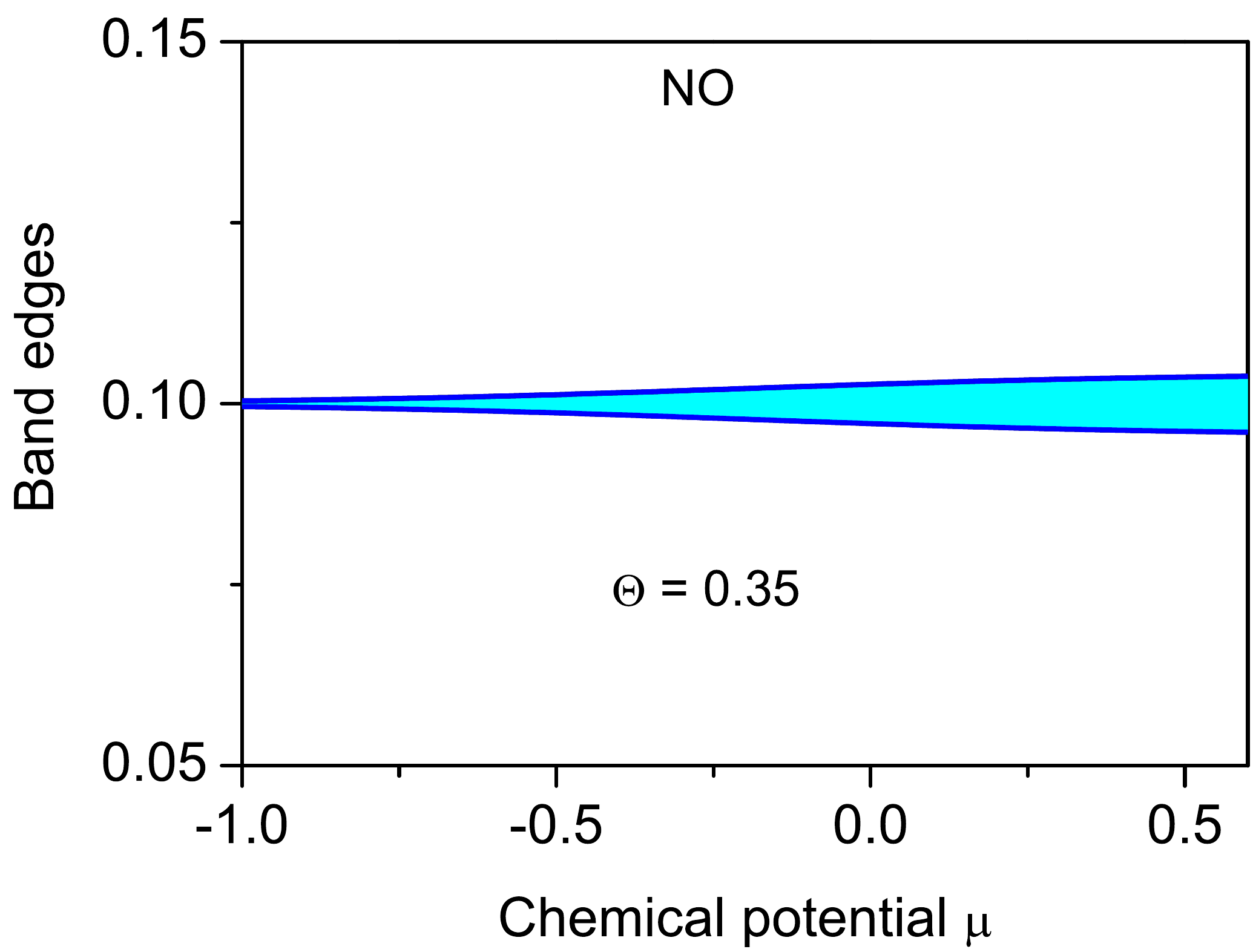}%
\caption{(Color online) Excitation energy bands as functions of boson chemical
potential at various temperatures ($\Theta=k_{\mathrm{B}}T$).
Parameter values: $\delta=0.1$, $W=0.03$. Here and in
figures~\protect\ref{fig02}--\protect\ref{fig05} energetic
quantities are given in units of $\abs{t'_0}$; thin solid (dotted)
vertical lines correspond to 1st (2nd) order PT.}
\label{fig01}
\end{figure}

\begin{figure}[!t]
\noindent%
\includegraphics[width=0.45\textwidth]{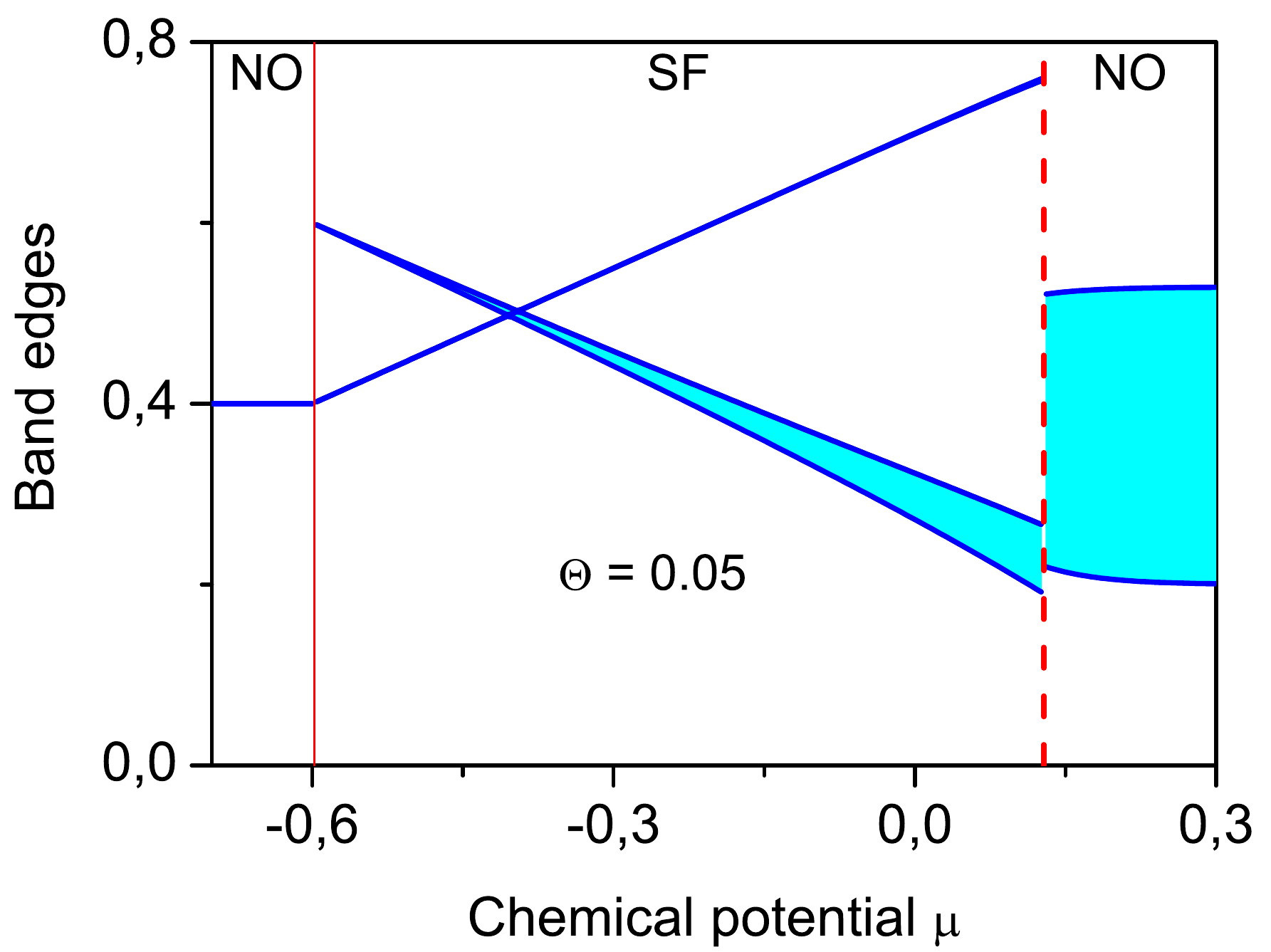}%
\hfill%
\includegraphics[width=0.45\textwidth]{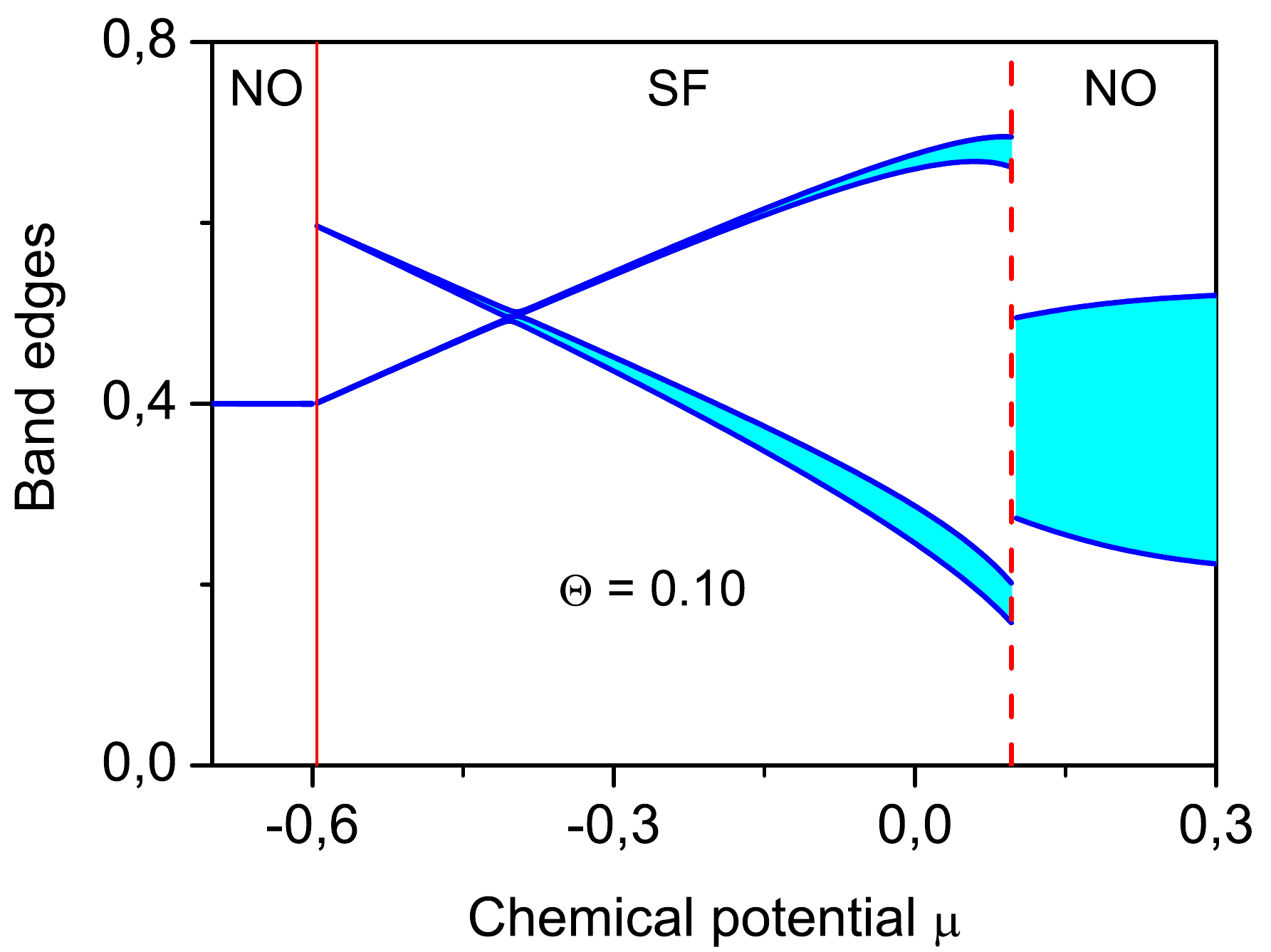}%
\\%
\smallskip
\noindent%
\includegraphics[width=0.45\textwidth]{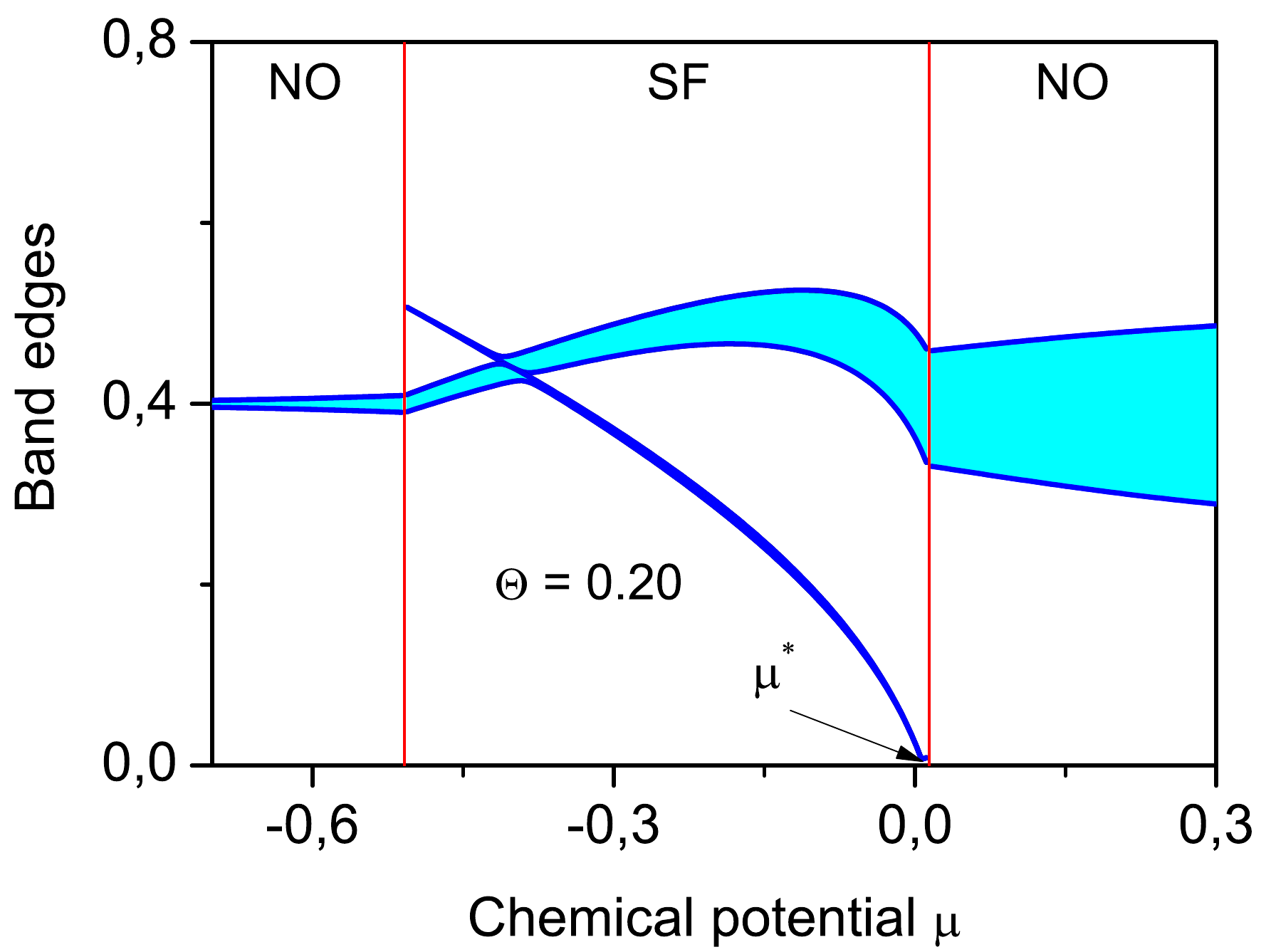}%
\hfill%
\includegraphics[width=0.455\textwidth]{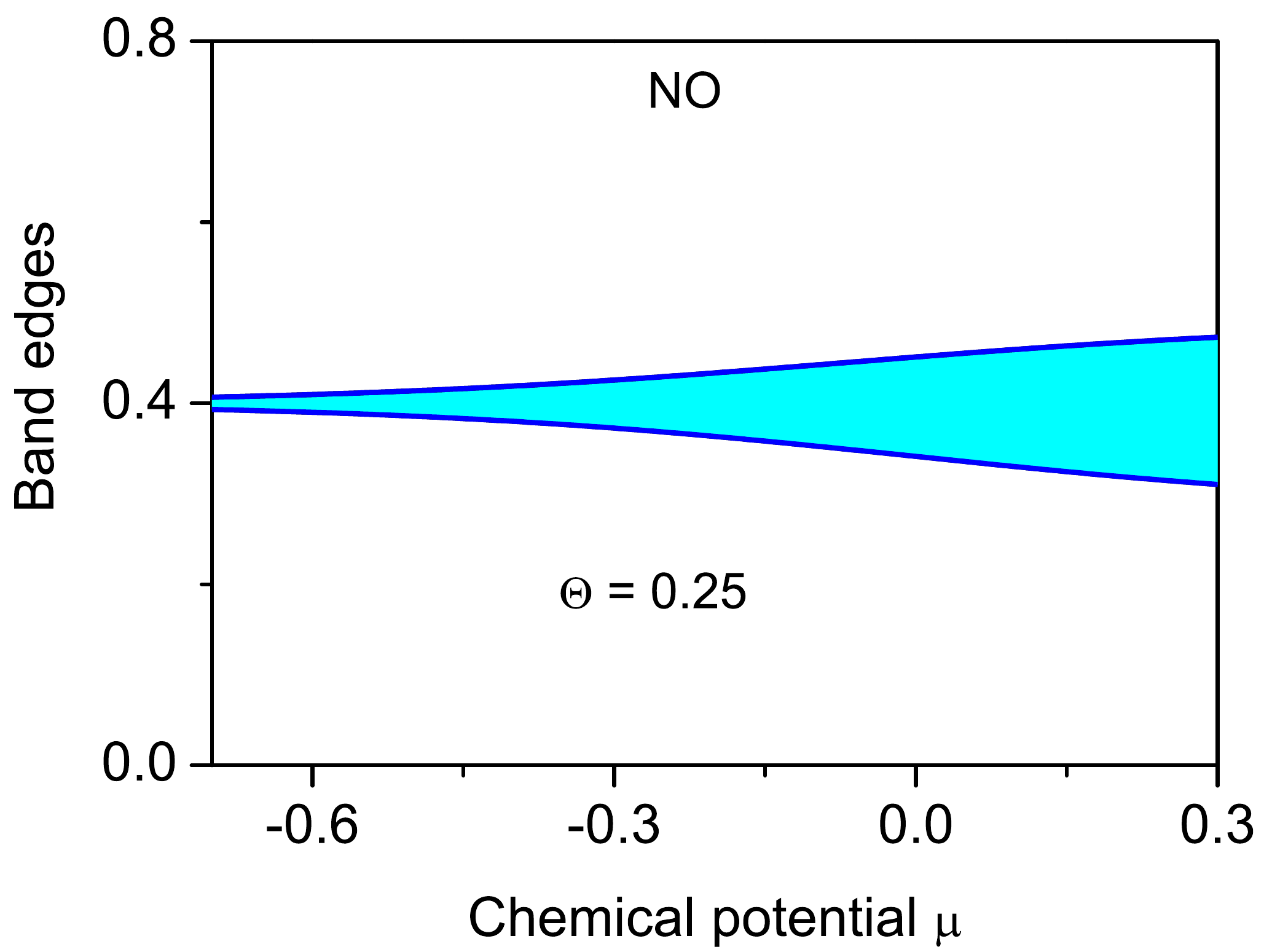}%
%

\caption{(Color online) Excitation energy bands as functions of boson chemical
potential at various temperatures. Parameter values: $\delta=0.4$,
$W=0.15$.}
\label{fig02}
\end{figure}

Band edges of the $\varepsilon_{1\ldots{4}}(\vec{q})$ branches can
be obtained putting $d^{2}\Phi_{q}=\pm W$ (a 2W parameter
determines the interval of the change of the Fourier-transform
$\Phi_{q}$ within the 1st Brillouin zone). Depending on the boson
chemical potential value, the excitation spectrum (\ref{eq3-21})
changes its form both in the range of NO or SF phase and at the
transition between them. This is illustrated in
figures~\ref{fig01} and \ref{fig02} where the position of band
edges as function of $\mu$ at various temperatures is shown (for
the sake of simplicity we present only positive branches of the
spectrum).

In normal phase (when BE-condensate is absent) $\alpha=0$,
$B_{q}=C_{q}=0$, the expression (\ref{eq3-18}) for the Green's
function greatly simplifies
\begin{equation}
\langle\langle\hat{x}|\hat{x}\rangle\rangle_{\omega,\vec{q}}^{\text{(NO)}}=-\frac{\hbar}{2\pi}d^{2}\frac{2\Delta_{21}\sigma}
{\hbar^{2}\omega^{2}-\Delta^{2}_{21}+2\sigma d^{2}\Phi_{q}\Delta_{21}}.
\label{eq3-23}
\end{equation}
Here, two branches remain, whose energies are equal in modulus but
differ among themselves by sign.
\begin{equation}
\varepsilon_{1,2}^{\text{(NO)}}=\pm\left[\Delta_{21}^{2}-2\sigma d^{2}\Phi_{q}\Delta_{21}\right]^{1/2}.
\label{eq3-24}
\end{equation}

Among the branches existing in the SF phase spectrum, two branches
($\varepsilon_{1}$ and $\varepsilon_{2}$) transform into branches
$\varepsilon_{1}^{\text{(NO)}}$ and
$\varepsilon_{2}^{\text{(NO)}}$ at the phase transition to the NO
phase. The other two ($\varepsilon_{3}$ and $\varepsilon_{4}$) are
new ones; they appear in the SF phase due to the presence of the
BE condensate. This can be seen while applying the expansion in
power series of the rotation angle $\alpha$ [see expressions
(\ref{eq2-11}) and (\ref{eq2-12})] at small values of the order
parameter $\xi$ (near the border of the SF phase region in the
case of SF-NO transition of the second order). In this case we
obtain:
\begin{align}
\varepsilon_{1,2}^{2}&=\Delta_{21}^{2}-2\sigma\Delta_{21}d^{2}\Phi_{q}(1-\alpha^{2})+\frac{4Q'\sigma\Delta_{01}\Delta_{21}d^{4}\Phi_{q}^{2}}
{\Delta_{21}^{2}-\Delta_{01}^{2}-2\sigma\Delta_{21}d^{2}\Phi_{q}}\alpha^{2},\nonumber\\
\varepsilon_{3,4}^{2}&=\Delta_{01}^{2}+\frac{2Q'\Delta_{01}(\Delta_{01}^{2}-\Delta_{21}^{2})d^{2}\Phi_{q}}{\Delta_{21}^{2}
-\Delta_{01}^{2}-2\sigma\Delta_{21}d^{2}\Phi_{q}}\alpha^{2}.
\label{eq3-25}
\end{align}
%

\begin{figure}[!t]
\centerline{
\includegraphics[width=0.5\textwidth]{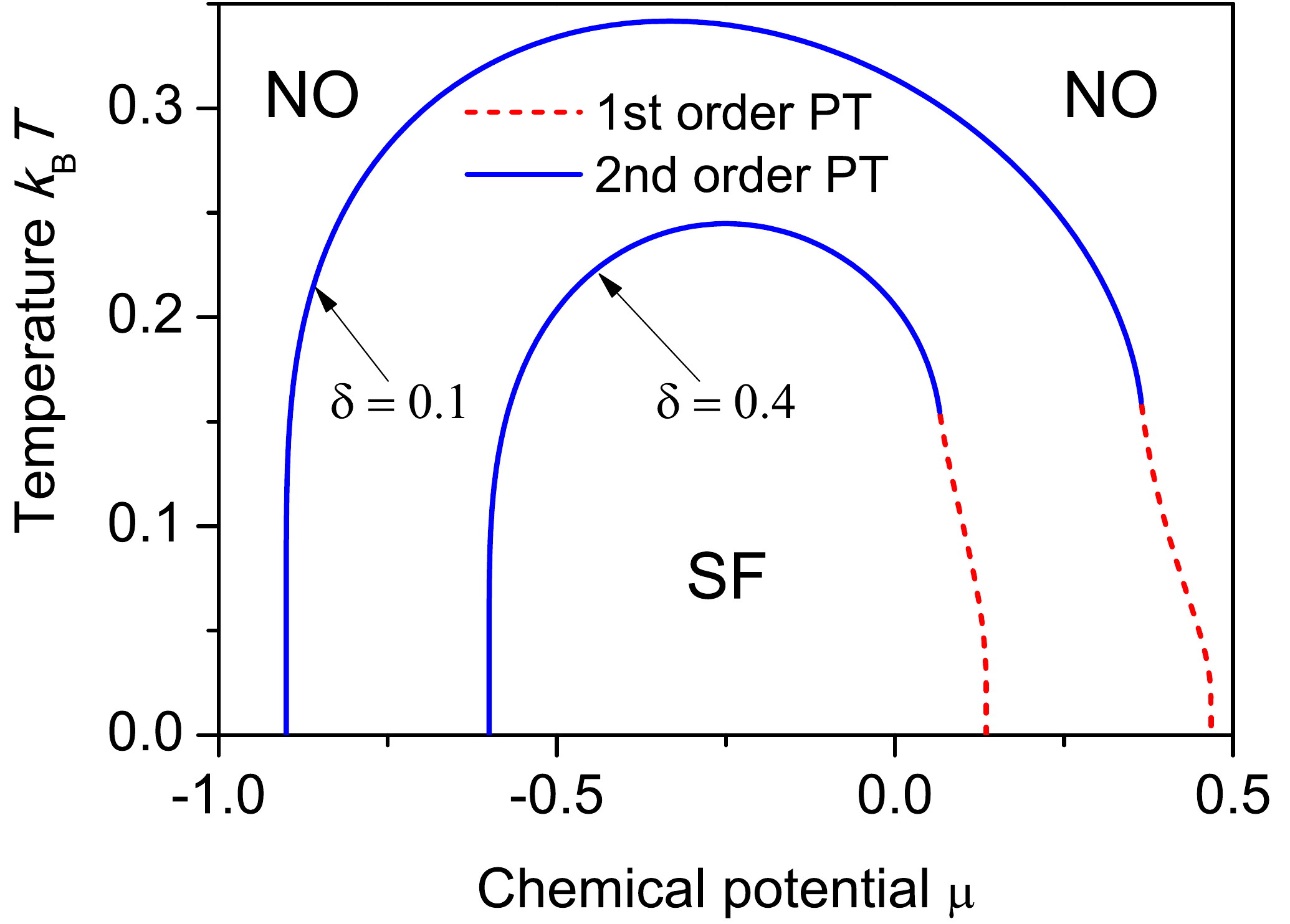}
}
\caption{(Color online) Mean-field phase diagram of the two-state Bose-Hubbard
model \protect\cite{Stasyuk24}.}
\label{fig03}
\end{figure}

At a small $\Phi_{q}$ interaction, when  the linear approximation
can be used,
\begin{equation}
\varepsilon_{1,2}=\pm\left[\Delta_{21}-\sigma d^{2}\Phi_{q}(1-\alpha^{2})\right],\qquad
\varepsilon_{3,4}=\pm\left[\Delta_{01}-Q' d^{2}\Phi_{q}\alpha^{2}\right].
\label{eq3-26}
\end{equation}
One can see that
$\varepsilon_{1,2}\rightarrow\varepsilon_{1,2}^{\text{(NO)}}$,
$\varepsilon_{3,4}\rightarrow\pm \Delta_{01}$ at
$\alpha\rightarrow0$. When passing  more  deeply into SF phase,
when $\xi$ parameter (angle $\alpha$) increases, the  widths  of
bands that correspond to the branches of spectrum are changed. The
bands $\varepsilon_{3}$ and $\varepsilon_{4}$ become broader while
the bands $\varepsilon_{1}$ and $\varepsilon_{2}$ become  narrower
(see figures~\ref{fig01} and \ref{fig02}).
At the point of the 1st order phase transition (see the phase
diagram, figure~\ref{fig03}), the positions of bands are changed
abruptly while in the case of the 2nd order PT, such a change is
continuous.
As a whole, the weights of the bands diminish when chemical
potential $\mu$ decreases and passes from a positive region to a
negative one.
Besides that, the redistribution of  their statistical weights
takes place. The  effect can be described by means of the
corresponding spectral densities. One  can  calculate the latter
using the Green's function
$\langle\langle\hat{x}|\hat{x}\rangle\rangle$.

Starting from (\ref{eq3-18}) we can  write the Green's function
$\langle\langle \hat{x}|\hat{x}\rangle\rangle$ in the following
form:
\begin{equation}
\langle\langle\hat{x}|\hat{x}\rangle\rangle_{\omega,q}=-\frac{\hbar}{2\pi}d^{2}\frac{Q(\hbar \omega)}
{(\hbar^{2}\omega^{2}-y_{1})(\hbar^{2}\omega^{2}-y_{2})},
\label{eq4-01}
\end{equation}
where
\begin{equation}
Q(\hbar \omega)=2\Delta_{01}\sin^{2}\alpha Q'\left(\hbar^{2}\omega^{2}-\Delta_{21}^{2}\right)
+2\Delta_{21}\cos^{2}\alpha\sigma\left(\hbar^{2}\omega^{2}-\Delta_{01}^{2}\right).
\label{eq4-02}
\end{equation}
After decomposition into simple fractions
\begin{align}
\langle\langle\hat{x}|\hat{x}\rangle\rangle_{\vec{q},\omega}&=-\frac{\hbar}{2\pi}d^{2}
\left[\frac{1}{\hbar \omega-\sqrt{y_{1}}}\;
\frac{Q(\sqrt{y_{1}})}{2\sqrt{y_{1}}(y_{1}-y_{2})}-\frac{1}{\hbar \omega +\sqrt{y_{1}}}\;\frac{Q(-\sqrt{y_{1}})}{2\sqrt{y_{1}}(y_{1}-y_{2})}\right.
\nonumber\\
&\left.\!\quad{}+\frac{1}{\hbar \omega -\sqrt{y_{2}}}\;\frac{Q(\sqrt{y_{2}})}{2\sqrt{y_{2}}(y_{2}-y_{1})}-\frac{1}{\hbar \omega+\sqrt{y_{2}}}\;
\frac{Q(-\sqrt{y_{2}})}{2\sqrt{y_{2}}(y_{2}-y_{1})}\right].
\label{eq4-03}
\end{align}
Since $Q(\sqrt{y_{i}})=Q(-\sqrt{y_{i}})$, the formula
(\ref{eq4-03}) can be written as follows:
\begin{align}
\langle\langle\hat{x}|\hat{x}\rangle\rangle_{\vec{q},\omega}&=-\frac{1}{2\pi}\frac{d^{2}}{2}\frac{1}{\varepsilon_{1}^{2}-\varepsilon_{3}^{2}}
\left[\frac{Q(\varepsilon_{1})}{\varepsilon_{1}}
\left(\frac{1}{\omega-\varepsilon_{1}/\hbar}-\frac{1}{\omega-\varepsilon_{2}/\hbar}\right)\right.
\nonumber\\
&\left.\!\quad{}-\frac{Q(\varepsilon_{3})}{\varepsilon_{3}}\left(\frac{1}{\omega-\varepsilon_{3}/\hbar}-\frac{1}{\omega-\varepsilon_{4}/\hbar}\right)\right].
\label{eq4-04}
\end{align}
Spectral density calculated per lattice site is given by the
relation:
\begin{equation}
\rho(\omega)=\frac{1}{N}\sum_{q}\rho_{q}(\omega),
\label{eq4-05}
\end{equation}
where
\begin{equation}
    \rho_{q}(w)
    =
    -2\Imp\langle\langle\hat{x}|\hat{x}\rangle\rangle_{\vec{q},\omega+\mathrm{i}\varepsilon}.
    \label{eq4-06}
\end{equation}
It is easy to determine the imaginary part of the function
(\ref{eq4-04}); then
\begin{align}
\rho_{q}(\omega)&=-\frac{d^{2}}{2(\varepsilon_{1}^{2}-\varepsilon_{3}^{2})}
\left\{\frac{Q(\varepsilon_{1})}{\varepsilon_{1}}\left[\delta (\omega-\varepsilon_{1}/\hbar)-\delta (\omega-\varepsilon_{2}/\hbar)\right]\right.\nonumber\\
&\left.\!\quad{}-\frac{Q(\varepsilon_{3})}{\varepsilon_{3}}\left[\delta(\omega-\varepsilon_{3}/\hbar)-\delta(\omega-\varepsilon_{4}/\hbar)\right]\right\}.
\label{eq4-07}
\end{align}

In the case of normal phase, the expression (\ref{eq4-07}) for
function $\rho_{q}(\omega)$ is more simple. Here, at $\alpha=0$,
\begin{equation}
Q(\hbar \omega)=2\Delta_{21}\sigma(\hbar^{2}\omega^{2}-\Delta_{01}^{2}),
\label{eq4-15}
\end{equation}
and, respectively,
\begin{equation}
Q(\varepsilon_{1})=2\Delta_{21}\sigma([\varepsilon_{1}^{\text{(NO)}}]^{2}-\Delta_{01}^{2}),\qquad
Q(\varepsilon_{3})=0.
\label{eq4-16}
\end{equation}
Thus, we have
\begin{equation}
\rho_{q}^{(N_{0})}(\omega)=-d^{2}\frac{\Delta_{21}\sigma}{\varepsilon_{1}^{\text{(NO)}}(q)}\left[\delta\left(\omega-\frac{\varepsilon_{1}^{\text{(NO)}}(q)}
{\hbar}\right)
-\delta\left(\omega-\frac{\varepsilon_{2}^{\text{(NO)}}(q)}{\hbar}\right)\right].
\label{eq4-17}
\end{equation}
The energies $\varepsilon_{1,2}^{\text{(NO)}}(\vec{q})$ are given
by formula (\ref{eq3-24}).

Additional branches, which appear in the spectrum in SF phase due
to BE condensate, possess an interesting feature. There exists a
possibility of nullification  of excitation energy at certain
values of wave vector when the chemical potential $\mu$ or the
temperature are changed [the
$\varepsilon_{3}(\vec{q})=-\varepsilon_{4}(\vec{q})$ function
tends in this case to zero at certain points on the border of the
1st Brillouin zone where $d^{2}\Phi_{q}=-W$, $W>0$]. At special
conditions, (see below) such a softening of the considered
vibrational mode could be present in the initial (NO) phase
serving as a manifestation of a tendency to instability with
respect to spatial modulation of the particle local displacements
as well as the BE condensate order parameter. We suppose, however,
that NO phase is stable in this sense; the relations between the
model parameters ensure the condition
\[
    \min\varepsilon_{1}^{\text{(NO)}}(q)
    >
    \left.\varepsilon_{q}^{\text{(NO)}}\right\vert_{d^{2}\Phi_{q}=-W}
    >
    0.
\]

Zero solution of equation (\ref{eq3-20}) exists  when the relation
\begin{equation}
\Delta_{21}^{2}\Delta_{01}^{2}\left[1-2d^{2}\Phi_{q}\left(\frac{\cos^{2}\alpha}{\Delta_{21}}\sigma+\frac{\sin^{2}\alpha}
{\Delta_{01}}Q'\right)\right]=0
\label{eq4-11a}
\end{equation}
fulfills. The energy $\varepsilon_{3}(\vec{q})$ goes to zero at
$\Delta_{01}=0$ (or $\Delta_{21}=0$), which corresponds to the
points $\mu^{*}$ indicated in figures~\ref{fig01} and \ref{fig02}
where the $\varepsilon_{i}(\vec{q})$ plots are presented, as well
as when the expression in square brackets  in (\ref{eq4-11a}) is
equal to zero. In the first of these cases, according to
(\ref{eq4-02}) and (\ref{eq4-04}), the statistical weight of this
branch also tends to zero. Due to that, the above mentioned
instability can be connected only with the  condition
\begin{equation}
1=2d^{2}\Phi_{q}\Psi,
\label{eq4-12a}
\end{equation}
where
\begin{equation}
\Psi=\frac{\cos^{2}\alpha}{\Delta_{21}}\sigma+\frac{\sin^{2}\alpha}{\Delta_{01}}Q'.
\label{eq4-12b}
\end{equation}

On the other hand, using this notation, we can write the Green's
function of displacements (\ref{eq3-18}) at zero frequency in such
a simple form
\begin{equation}
\langle\langle \hat{x}|\hat{x}
\rangle\rangle_{\omega=0,q}=-\frac{\hbar}{2\pi}d^{2}\frac{\Psi(\mu,T)}{1-2d^{2}\Phi_{q}\Psi(\mu,T)}. \label{eq4-13a}
\end{equation}
This function determines a  static susceptibility with respect to
the field which evokes the modulation characterized by the wave
vectors $\vec{q}$. Its divergence that takes place at the
(\ref{eq4-12a}) condition, is just a manifestation of the above
mentioned instability. The boundary of region where such an
instability exists is described by equation
\begin{equation}
1+2W \Psi(\mu,T)=0.
\label{eq4-14a}
\end{equation}

In the case of NO phase, this equation becomes simpler and takes
the following form:
\begin{equation}
1+2W\sigma/\Delta_{21}=0.
\label{eq4-15a}
\end{equation}
The NO phase is stable when
$1+2W\sigma/\Delta_{21}\equiv1+2W\langle
X^{00}-X^{11}\rangle/\Delta_{21}>0$.
In the $T=0$ limit and in the $\mu>0$ region, since $\langle
X^{rr}\rangle=\delta_{r1}$, this leads to the condition
\begin{equation}
\Delta_{21}\vert_{\text{(NO)}}\equiv\delta>2W
\label{eq4-16a}
\end{equation}
(in the case $\mu<0$ and at $T=0$, there are no bosons outside the
SF phase region  \cite{Micnas27a}).

\begin{figure}
\noindent%
\includegraphics[width=0.44\textwidth]{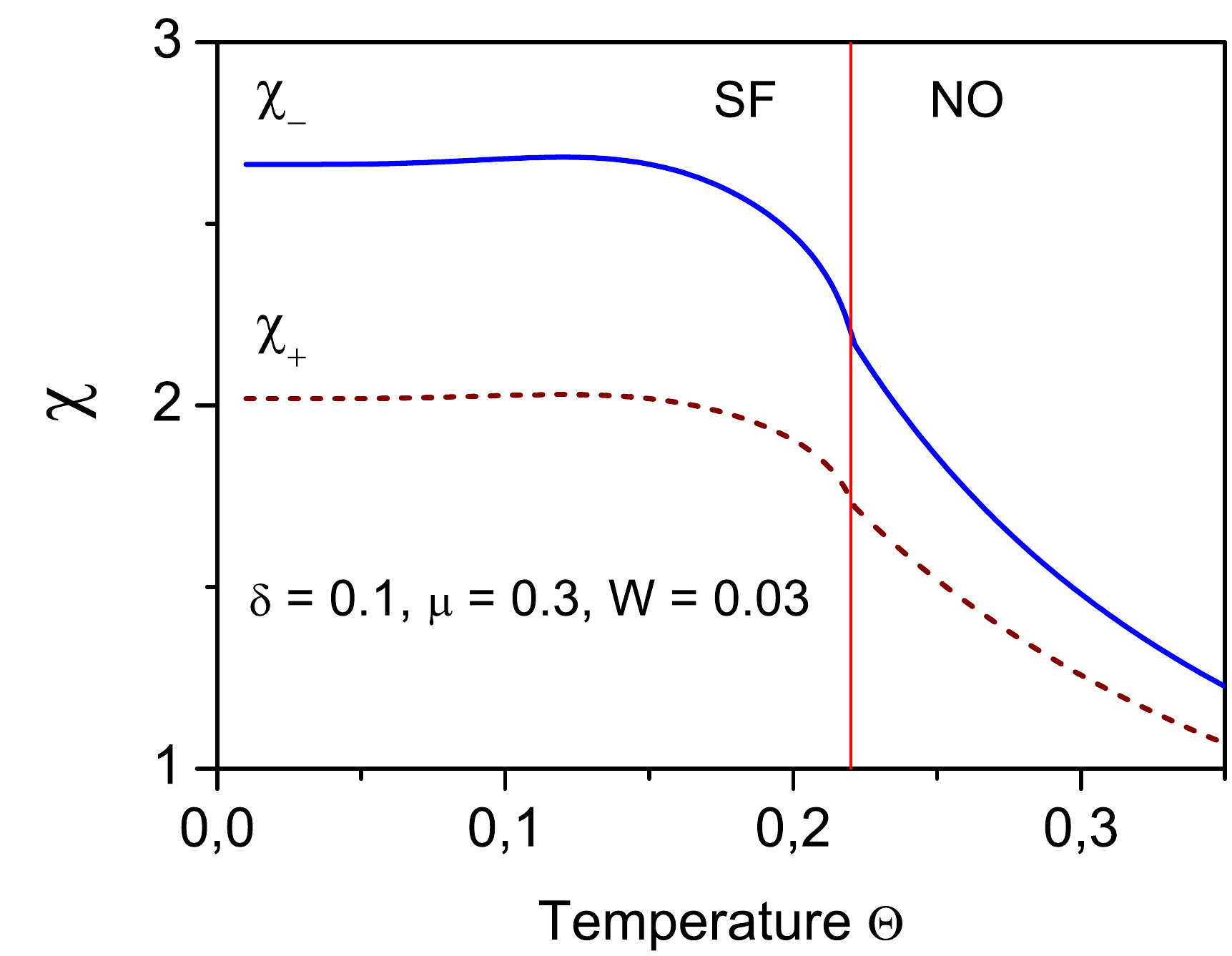}%
\hfill%
\includegraphics[width=0.45\textwidth]{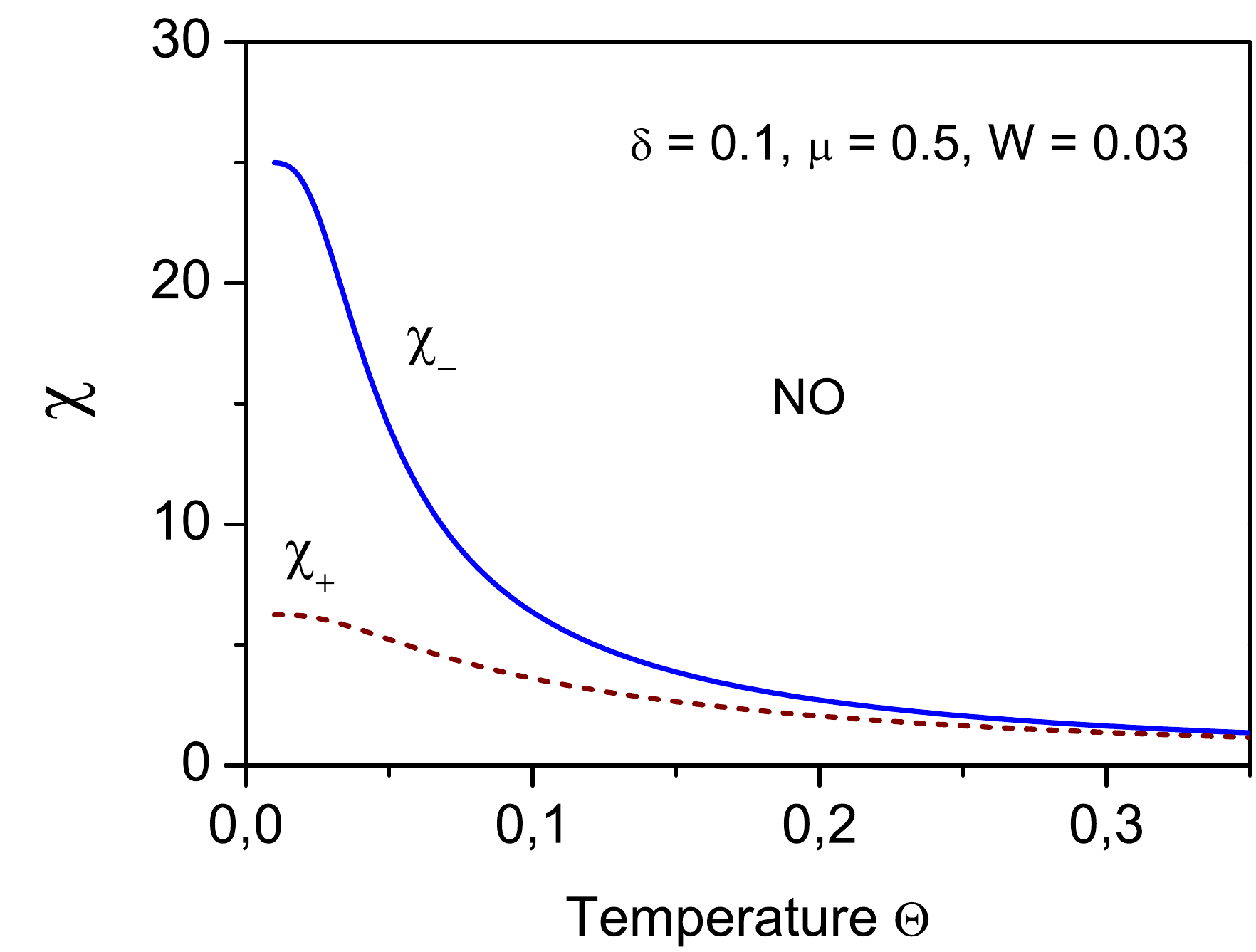}%
\\%
\smallskip
%
\includegraphics[width=0.45\textwidth]{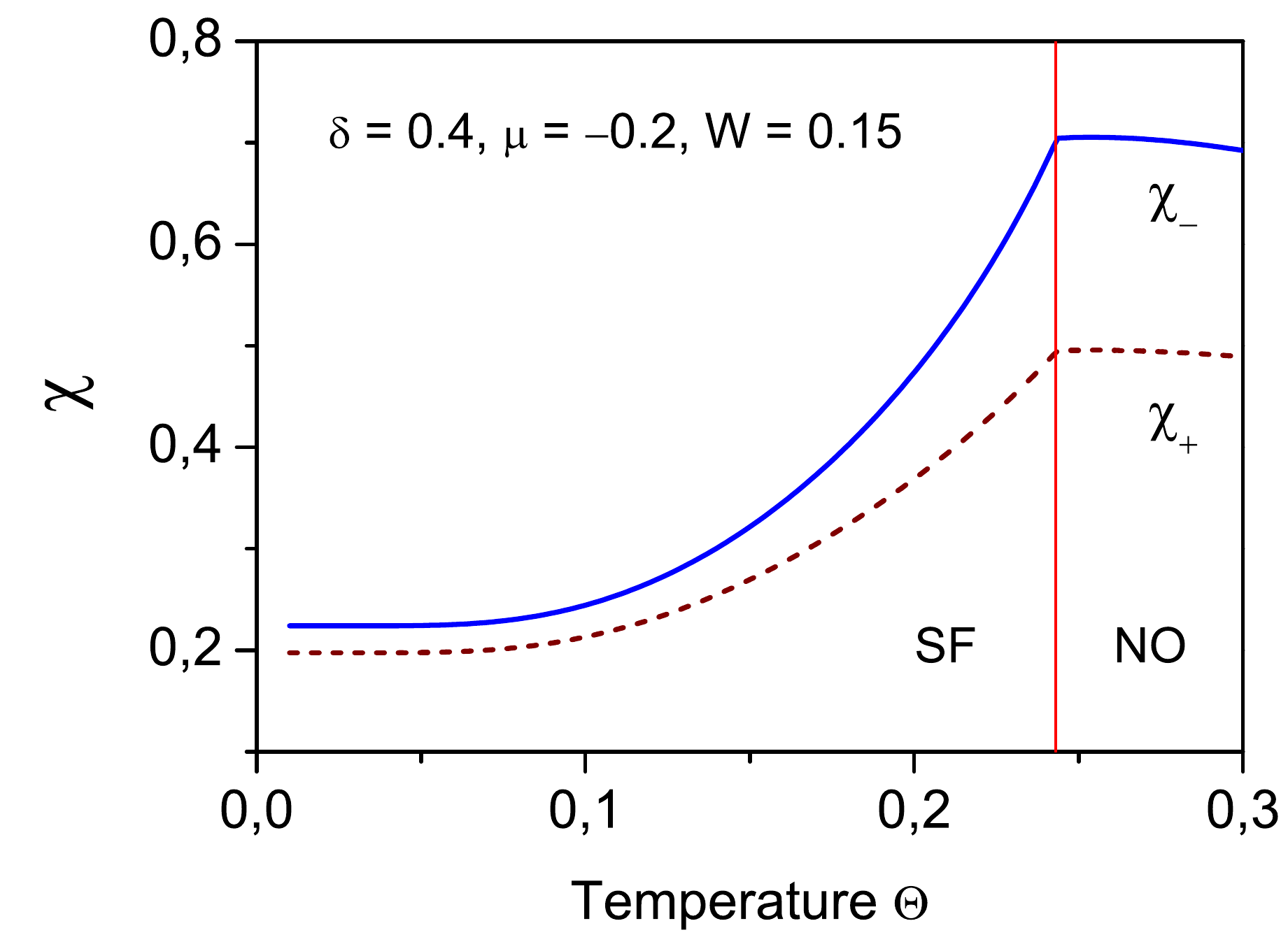}%
\hfill%
\includegraphics[width=0.45\textwidth]{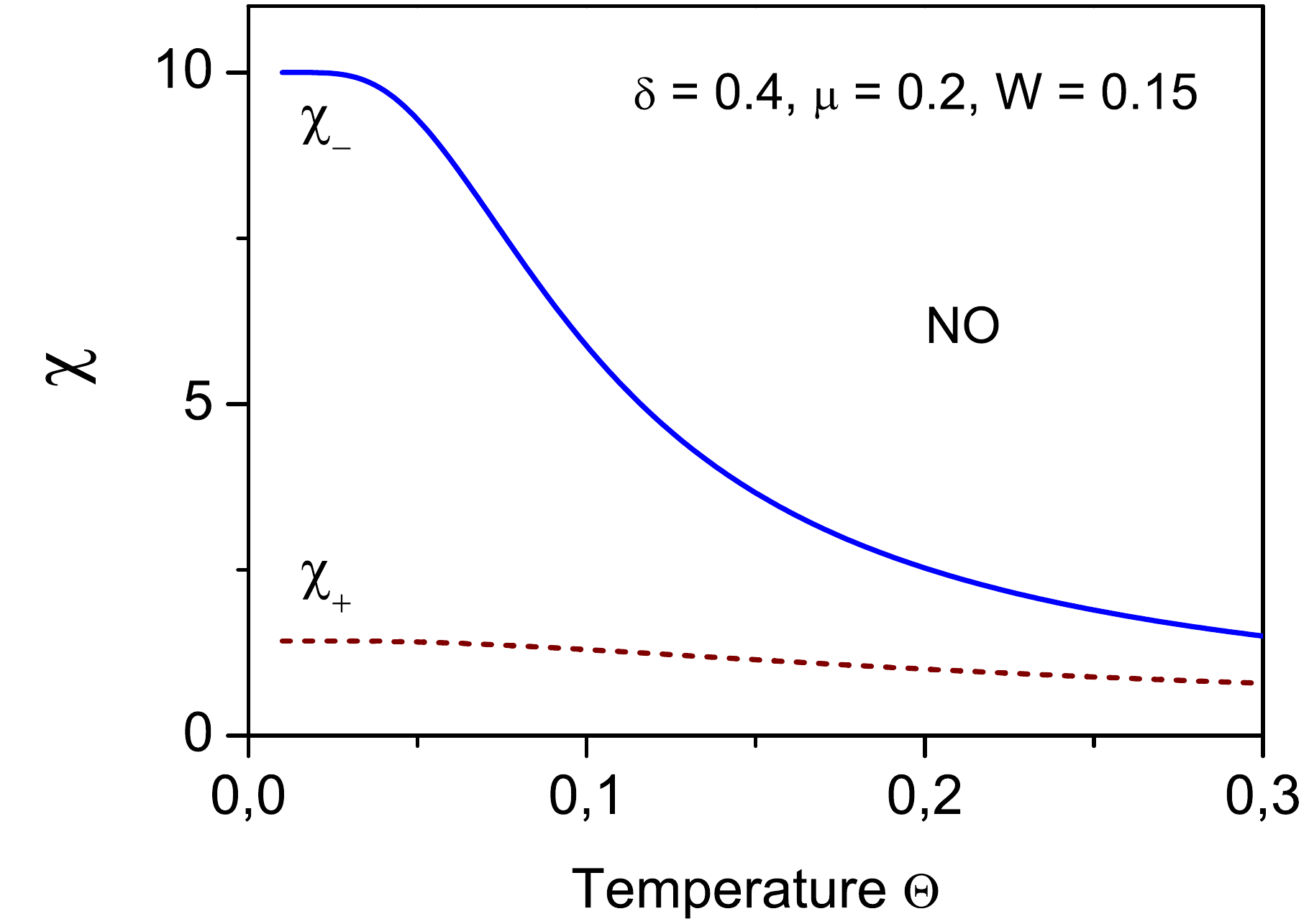}%
%
%
\caption{(Color online) Susceptibilities $\chi_{+}$ and $\chi_{-}$ as functions
of temperature at various values of boson chemical potential.}
\label{fig04}
\end{figure}

\begin{figure}[!t]
\centerline{
\includegraphics[width=0.5\textwidth]{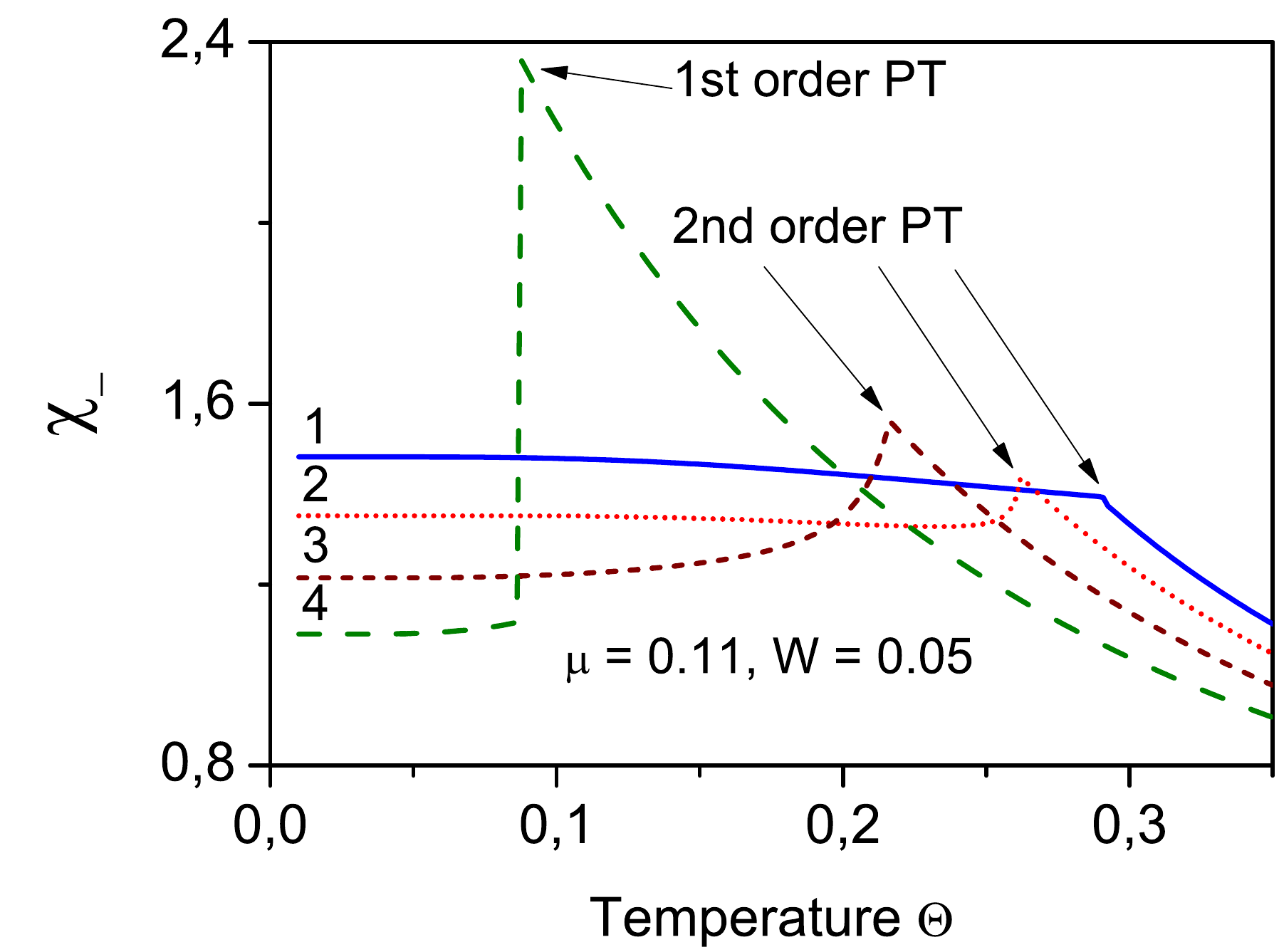}
\includegraphics[width=0.48\textwidth]{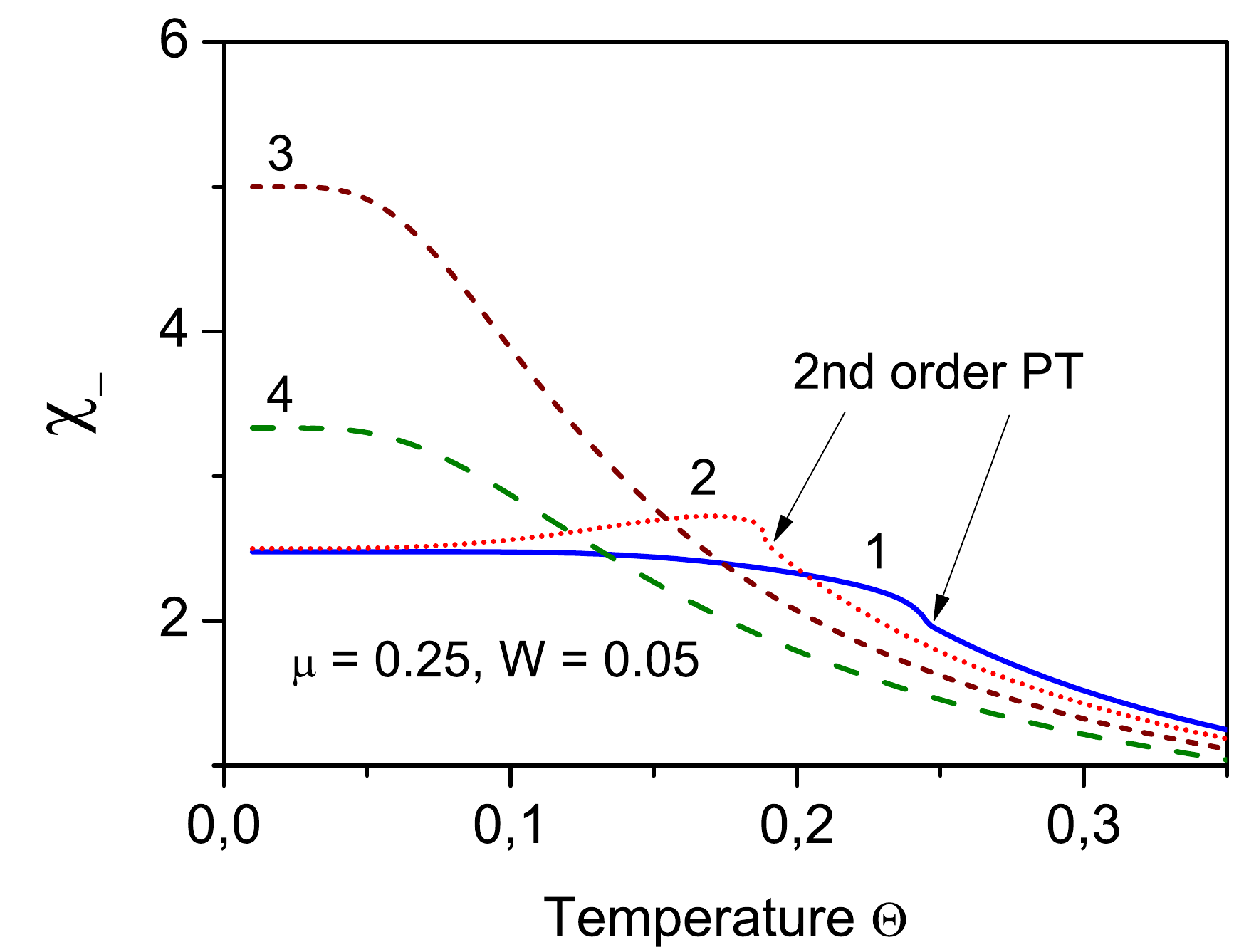}}%
\bigskip
\centerline{\includegraphics[width=0.5\textwidth]{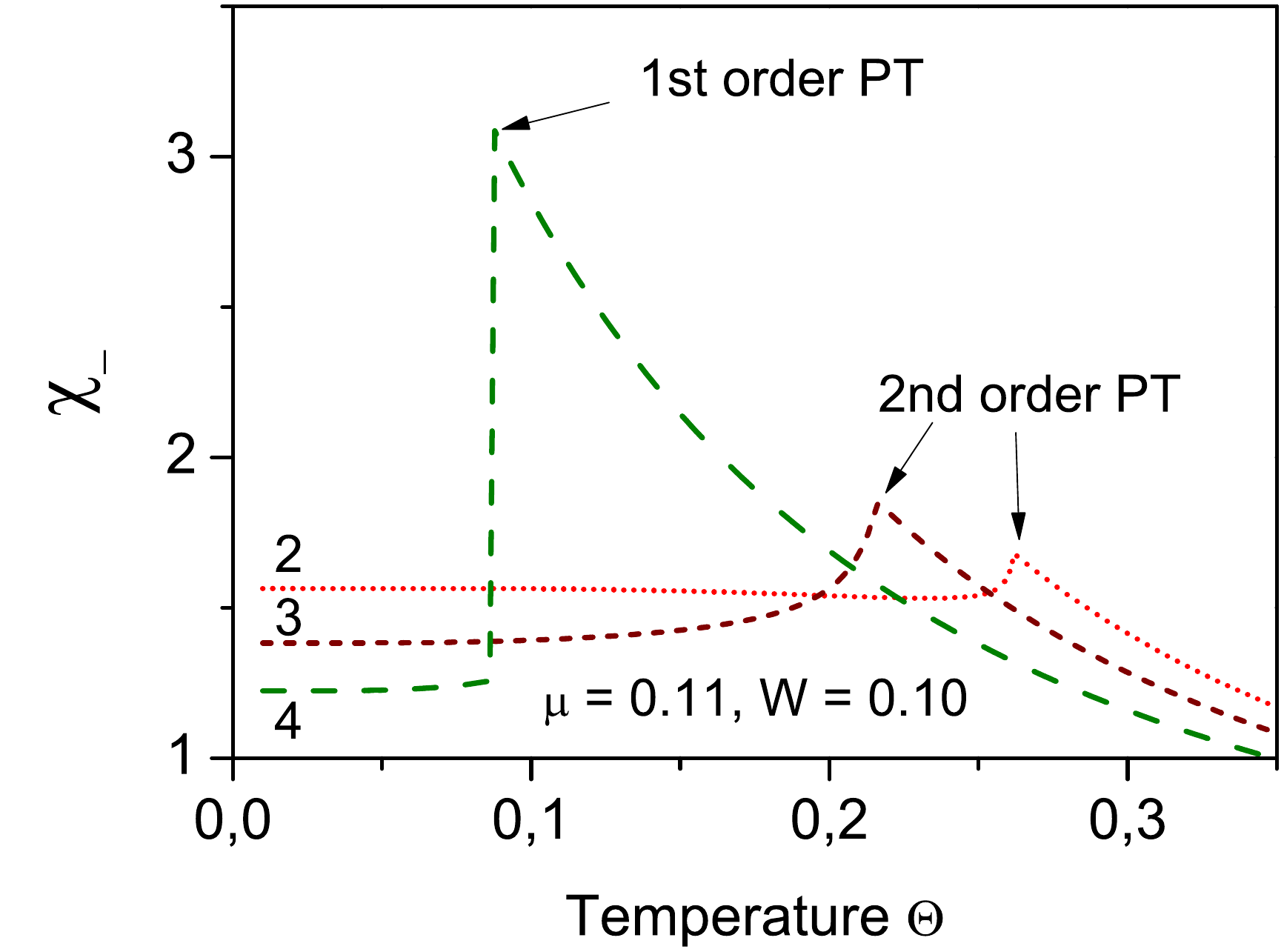}}%
\caption{(Color online) Susceptibility $\chi_{-}$ as function of temperature in
the region of NO-SF phase transition at different values of model
parameters.}
\label{fig05}
\end{figure}

In figures~\ref{fig04} and \ref{fig05} we show the graphs for
susceptibility
\begin{equation}
\chi_{\pm}=\left.-\frac{2\pi}{\hbar^{2}}\langle\langle\hat{x}|\hat{x}\rangle\rangle_{\omega=0,\vec{q}}\right\vert_{d^{2}\Phi_{q}=\pm
W} \label{eq3-43}
\end{equation}
as a function of temperature. The calculations were performed at
various values of $\mu$, $\delta$ and $W$ meeting the condition
$W<\delta/2$. As we can see, $\chi_{-}$ is greater than
$\chi_{+}$; this gives an  evidence for the dominance of potential
instability with respect to modulation of particle displacements
(with the wave vector at the boundary of Brillouin zone).

Susceptibility $\chi_{-}$ increases noticeably at a lowering of
temperature in NO phase. However, at a further decrease of $T$
after transition to SF phase, a significant suppression of
$\chi_{-}$ takes place. The $\chi_{-}$ function (just as
$\chi_{+}$) approaches saturation (increasing or decreasing
slightly, depending on value of $\delta$). In the majority of
cases, there exists a  peak of the $\chi_{-}$ function in the
vicinity of the 1st order PT point. At this point, $\chi_{-}$
reaches a maximum value which remains finite in the considered
range of the model parameter values. Consequently, the tendency to
the displacement modulation, that is characteristic of NO phase,
weakens when the NO-SF transition takes place and the BE
condensate appears.

\section{Conclusions}

In this work, for the system of quantum particles in a lattice
(Bose-atoms in optical lattice, adsorbed or intercalated particles
in a crystal), we performed an investigation of the spectrum of
collective vibrational excitations that appear at the
displacements of a particle with respect to their equilibrium
positions in local positions and which are connected with
interactions between them. At a  quantum description of
displacements, we took into account the transitions of particles
between the ground state and the first excited state only.

Such excitations, by their nature, are analogous to optical
phonons in a usual crystal lattice. On the other hand, in the
two-level approximation, they are similar to the Frenkel excitons
or to excitations in the systems with double potential wells.

Calculation of the spectrum is performed in the framework of such
a two-state model supposing that the quantum hopping of  particles
between the neighbouring positions in a lattice takes place in
their excited state. Consideration is performed on the basis of
the hard-core boson model.

We found the excitation spectrum in the random phase approximation
using the Green's function method. It is shown that the spectrum
is of a band character and consists of positive and negative parts
(among them, the second part is the mirror image of the first
one). Only one positive branch [$\varepsilon_{1}(\vec{q})$] exists
in the NO phase, while in the SF phase there appears an additional
branch [$\varepsilon_{3}(\vec{q})$].

Positions, spectral weights of  branches and widths of the
corresponding bands change at the shift of the chemical potential
of bosons and depend on temperature. A noticeable redistribution
of the intensities of  branches takes place in  the SF phase at
the change of $\mu$.

An additional branch [$\varepsilon_{3}(\vec{q})$] can become
``soft'' and go to zero at a certain value of $\mu$. However, such
a behaviour does not lead to an instability in the system with
respect to the modulation of particle displacements. The spectral
weight of this mode also diminishes in this case and is equal to
zero at this point. This fact is confirmed by calculation of the
related susceptibility $\chi_{-}$ connected directly with Green's
function of displacements. The results show that in the case, when
the boson system is stable in the NO phase, the transition to the
SF phase and the subsequent lowering  of $T$ (or  decrease of
$\mu$) do not give rise to instability.
The possible displacement modulation could lead to the appearance
of the so-called super-solid phase (with modulation of  the BE
condensate order parameter). In  our case, however, the BE
condensate remains uniform.

The analysis of the spectrum of phonon-like excitations in optical
lattices and the investigation of its transformation and the
appearance of new branches in the SF phase are possible with the
use of Raman spectroscopy.
Raman scattering intensity can be expressed in this case in terms
of the Green's function of particle displacements of the
$\langle\langle\hat{x}|\hat{x}\rangle\rangle$
type, and the poles of this function will determine the Raman
excitation spectrum (additional poles, appearing in the SF phase,
manifest themselves as new Raman lines). Experimentally, such a
technique was applied in \cite{Mueller06} to study the spectrum of
ultracold Bose-atoms excited to the upper Bloch band. Attention
was paid to the observed features of scattering line profiles in a
MI phase. Their explanation was given in terms of local
transitions of bosons between ground and excited states in
potential wells in a lattice.
Theoretical analysis of Raman scattering intensity due to
transitions to higher vibrational bands (in the case of
sufficiently deep lattice and different number of particles per
site) was performed in \cite{Blakie06}, as well as for the MI
regime. The attention to differences between the intensity or
amplitude of scattered light for SF and MI phases was directed in
\cite{Zhou10}, where the possible experiments based on the Raman
scattering-in-cavity technique were proposed. It is clear, on the
whole, that Raman spectroscopy opens up new possibilities in the
study of quantum states and the dynamics of Bose-atoms in the
presence of BE condensate. However, more systematic investigations
in this direction are needed. In this connection, a further
development of the theory based on the calculation of the
effective Raman coupling strength for specific cases and models is
necessary.

\ukrainianpart

\title{╟сєфцхээ  Їюэюээюую Єшяє є фтюёЄрэют│щ ьюфхы│ ┴ючх-╒рссрЁфр}
\author{▓.┬. ╤Єрё■ъ, ╬.┬. ┬хышўъю, ╬. ┬юЁюсщют}
\address{▓эёЄшЄєЄ Ї│чшъш ъюэфхэёютрэшї ёшёЄхь ═└═ ╙ъЁр┐эш,
тєы. ▓.~╤т║эЎ│Ў№ъюую,~1, 79011 ╦№т│т, ╙ъЁр┐эр}

\makeukrtitle

\begin{abstract}
\tolerance=3000%
─юёы│фцє║Є№ё  ёяхъЄЁ ъюыхъЄштэшї чсєфцхэ№ Їюэюээюую Єшяє т ёшёЄхь│
сючх-рЄюь│т є юяЄшўэ│щ ┤ЁрЄЎ│ (с│ы№° чруры№эю, є ёшёЄхь│ ътрэЄютшї
ўрёЄшэюъ,  ъ│ юяшёє■Є№ё  ьюфхыы■ ┴ючх-╒рссрЁфр). ╥ръ│ чсєфцхээ 
тшэшър■Є№ чртф ъш чь│∙хээ ь ўрёЄшэюъ т│фэюёэю ┐ї ыюъры№эшї
Ё│тэютрцэшї яючшЎ│щ. ┬шъюЁшёЄрэю фтюЁ│тэхтє ьюфхы№,  ър яЁшщьр║ фю
єтруш яхЁхїюфш сючюэ│т ь│ц юёэютэшь │ яхЁ°шь чсєфцхэшь ёЄрэюь є
яюЄхэЎ│ры№эшї  ьрї, р Єръюц тчр║ьюф│■ ь│ц эшьш. ╨ючЁрїєэъш
яЁютхфхэю є эрсышцхээ│ їрюЄшўэшї Їрч Єр т уЁрэшЎ│ цюЁёЄъшї
сючюэ│т. ╧юърчрэю, ∙ю ёяхъЄЁ чсєфцхэ№ ёъырфр║Є№ё  є эюЁьры№э│щ
Їрч│ ч юфэ│║┐ чюэш хъёшЄюээюую Єшяє, т Єющ ўрё  ъ є Їрч│ ч
сючх-ъюэфхэёрЄюь тшэшър║ фюфрЄъютр чюэр. ╨ючЄр°єтрээ , ёяхъЄЁры№э│
труш Єр °шЁшэш чюэ ёєЄЄ║тшь ўшэюь чрыхцрЄ№ т│ф ї│ь│ўэюую
яюЄхэЎ│рыє сючюэ│т Єр ЄхьяхЁрЄєЁш. ╬суютюЁ■■Є№ё  єьютш ёЄ│щъюёЄ│
ёшёЄхьш т│фэюёэю чэшцхээ  ёшьхЄЁ│┐ Єр ьюфєы Ў│┐ чь│∙хэ№.
\keywords ьюфхы№ ┴ючх-╒рссрЁфр, цюЁёЄъ│ сючюэш, сючх-ъюэфхэёрЎ│ ,
чсєфцхэр чюэр, Їюэюэш
\end{abstract}


\begin{thebibliography}{99}
%
\bibitem{wrk01} Fisher M.P.A., Weichman P.B., Grinstein G., Fisher D.S.,
    Phys. Rev. B, 1989, \textbf{40}, 546;
    \bibdoi{10.1103/PhysRevB.40.546}.

\bibitem{Jaksch02} Jaksch D., Bruder C., Cirac J.I.,
    Gardiner~C.W., Zoller~P., Phys. Rev. Lett., 1998, \textbf{81},
    3108;\\
    \bibdoi{10.1103/PhysRevLett.81.3108}.

\bibitem{wrk04} Zwerger W., J. Opt. B: Quantum Semiclass. Opt., 2003,
    \textbf{5}, S9; \bibdoi{10.1088/1464-4266/5/2/352}.


\bibitem{wrk05} Damski B., Zakrzewski J., Santos L., Zoller P., Lewenstein
    M., Phys. Rev. Lett., 2003, \textbf{91}, 080403; \\
    \bibdoi{10.1103/PhysRevLett.91.080403}.


\bibitem{wrk06} Konabe S., Nikuni T., Nakamura M., Phys. Rev. A, 2006,
    \textbf{73}, 033617;
    \bibdoi{10.1103/PhysRevA.73.033617}.

\bibitem{wrk03} Dupuis N., Sengupta K., Physica B, 2009,
    \textbf{404}, 517; \bibdoi{10.1016/j.physb.2008.11.084}.

\bibitem{Astaldi07} Astaldi C., Bianco A., Modesti S., Tosatti E., Phys.
    Rev. Lett., 1992, \textbf{68}, 90;
    \bibdoi{10.1103/PhysRevLett.68.90}.

\bibitem{Nishijima08} Nishijima M., Okuyama H., Takagi N., Aruga T., Brenig
    W., Surf. Sci. Rep., 2005, \textbf{57}, 113; \\
    \bibdoi{10.1016/j.surfrep.2005.03.001}.

\bibitem{Reilly09} Reilly P.D., Harris R.A., Whaley K.B., J. Chem. Phys.,
    1991, \textbf{95}, 8599; \bibdoi{10.1063/1.461239}.

\bibitem{Ignatyuk10} Ignatyuk V.V., Phys. Rev. E, 2009,
    \textbf{80}, 041133; \bibdoi{10.1103/PhysRevE.80.041133}.

\bibitem{Velychko11} Velychko O.V., Stasyuk I.V., Condens. Matter Phys.,
    2009, \textbf{12}, 249; \bibdoi{10.5488/CMP.12.2.239}.

\bibitem{Mysakovych12} Mysakovych T.S., Krasnov V.O., Stasyuk I.V., Ukr. J.
    Phys., 2010, \textbf{55}, 228.

\bibitem{Puska13} Puska M.J., Nieminen R.M., Surf. Sci., 1985,
    \textbf{157}, 413; \bibdoi{10.1016/0039-6028(85)90683-1}.

\bibitem{Brenig14} Brenig W., Surf. Sci., 1993, \textbf{291},
    207; \bibdoi{10.1016/0039-6028(93)91492-8}.

\bibitem{Muller15} M\"{u}ller T., F\"{o}lling S., Widera A., Bloch I.,
    Phys. Rev. Lett., 2007, \textbf{99}, 200405;
    \bibdoi{10.1103/PhysRevLett.99.200405}.

\bibitem{Scarola16} Scarola V.W., Das Sarma S., Phys. Rev. Lett., 2005,
    \textbf{95}, 033003; \bibdoi{10.1103/PhysRevLett.95.033003}.

\bibitem{Isacsson17} Isacsson A., Girvin S.M., Phys. Rev. A, 2005,
    \textbf{72}, 053604; \bibdoi{10.1103/PhysRevA.72.053604}.

\bibitem{Cai18} Cai Z., Wu C., Phys. Rev. A, 2011, \textbf{84},
    033635; \bibdoi{10.1103/PhysRevA.84.033635}.

\bibitem{Liu19} Liu B., Yu X.-L., Liu W.-M., Phys. Rev. A, 2013,
    \textbf{88}, 063605; \bibdoi{10.1103/PhysRevA.88.063605}.

\bibitem{Stojanovii20} Stojanovi\v{c} V.M., Wu C., Lin W.V., Das Sarma S.,
    Phys. Rev. Lett., 2008, \textbf{101}, 125301;\\
    \bibdoi{10.1103/PhysRevLett.101.125301}.

\bibitem{Kimura21} Kimura T., Tsuchiya S., Kurihara S., Phys. Rev. Lett.,
    2005, \textbf{94}, 110403;
    \bibdoi{10.1103/PhysRevLett.94.110403}.

\bibitem{Pai22} Pai R.V., Sheshadri K., Pandit R., Phys. Rev. B, 2008,
    \textbf{77}, 014503; \bibdoi{10.1103/PhysRevB.77.014503}.

\bibitem{Chen23} Chen G.-H., Wu Y.-S., Phys. Rev. A, 2003, \textbf{67},
    013606; \bibdoi{10.1103/PhysRevA.67.013606}.

\bibitem{Stasyuk24} Stasyuk I.V., Velychko O.V., Condens. Matter Phys., 2011,
    \textbf{14}, 13004; \bibdoi{10.5488/CMP.14.13004}.

\bibitem{wrk25} Stasyuk I., Velychko O., Theor. Math. Phys., 2011,
    \textbf{168}, 1347; \bibdoi{10.1007/s11232-011-0110-2}.

\bibitem{wrk26} Stasyuk I.V., Velychko O.V., Condens. Matter Phys., 2012,
    \textbf{15}, 33002; \bibdoi{10.5488/CMP.15.33002}.

\bibitem{Micnas27a} Micnas R., Ranninger J., Robaszkiewicz S., Rev. Mod.
    Phys., 1990, \textbf{62}, 113;
    \bibdoi{10.1103/RevModPhys.62.113}.

\bibitem{Mahan28} Mahan G.D., Phys. Rev. B, 1976, \textbf{14},
    780; \bibdoi{10.1103/PhysRevB.14.780}.

\bibitem{Stasyuk29} Stasyuk I.V., Dulepa I.R., J. Phys. Stud., 2009,
    \textbf{13}, 2701 (in Ukrainian).

\bibitem{Sengupta30} Sengupta K., Dupuis N., Majumdar P., Phys.
    Rev. A, 2007, \textbf{75}, 063625;
    \bibdoi{10.1103/PhysRevA.75.063625}.

\bibitem{wrk31} Hen I., Iskin M., Rigol M., Phys. Rev. B, 2010,
    \textbf{81}, 064503; \bibdoi{10.1103/PhysRevB.81.064503}.

\bibitem{Stasyuk32} Stasyuk I.V., Vorobyov O., Condens. Matter
    Phys., 2013, \textbf{16}, 23005; \bibdoi{10.5488/CMP.16.23005}.

\bibitem{Mueller06} M\"{u}ller T., Exploring excited Bloch bands in optical
    lattices via stimulated Raman transitions. Dip\-lom\-ar\-beit,
    Mainz University, 2006.

\bibitem{Blakie06} Blakie~P.B., New J. Phys., 2006, \textbf{8},
    157; \bibdoi{10.1088/1367-2630/8/8/157}.

\bibitem{Zhou10} Zhou~X., Xu~X., Yin~L., Liu~W.M.,
    Chen~X., Opt. Express, 2010, \textbf{18},
    15664; \bibdoi{10.1364/OE.18.015664}.
%
\end{thebibliography}
\end{document}